# Observation of Infrared and Radio Lines of Molecules toward GL 2591 and Comparison to Physical and Chemical Models

John S. Carr[1,2]
Department of Astronomy, The Ohio State University, Columbus, Ohio 43210–1106

and

Neal J. Evans II[2,3], and J. H. Lacy[2,4]
Department of Astronomy, The University of Texas at Austin, Austin, Texas 78712–1083

and

Shudong Zhou[5]
Astronomy Department, University of Illinois, Urbana, Illinois 61801 and
Institute of Astronomy & Astrophysics, Academia Sinica, P.O. Box 1-87, Nankang, Taipei, Taiwan 115, R.O.C.

## ABSTRACT

We have observed rovibrational transitions of $C_2H_2$ and HCN near 13 $\mu$m in absorption against GL2591. We have marginally detected $NH_3$ and set upper limits on rovibrational lines of $CH_4$, CS, SO, and SiO. We have also observed rotational transitions at 0.6–3 mm of CS, HCN, $H_2CO$, and $HCO^+$. The rovibrational data were analyzed in comparison to the absorption line analysis of CO by Mitchell et al. (1989). Our data are consistent with the $C_2H_2$ and HCN absorption arising in the same warm (200 K) and hot (1010 K) components seen in CO, but we see little evidence for the cold (38 K) component seen in CO. The results can be explained by a model in which early-time gas-phase abundances are preserved on grain mantles and later released at high temperature. Analysis of the rotational lines indicates that these do not arise from the same gas as the rovibrational lines. Comparison of the two data sets shows that the rovibrational absorption of HCN must come from a region with a small angular extent (less than about 2–3″, or about 2000–3000 AU at a distance of 1 kpc) and a much higher (factor of 400) abundance. The rovibrational lines from higher J states (J about 20) indicate that the hot HCN deviates from LTE. A good fit is obtained for a density of about $3 \times 10^7$ cm$^{-3}$. Analysis of the rotational lines, which arise in the extended cloud around the source, shows that no single-density model can explain all the data. Models with density and temperature gradients do much better; in particular models with $n(r) \propto r^{-\alpha}$, with $\alpha = 1.5$, can reproduce the observed pattern of emission line strengths. Models with $\alpha = 1.0$ or 2.0 are less satisfactory. These models predict densities of $3 \times 10^7$ cm$^{-3}$ at radii slightly smaller than, but similar to, the upper limits on the size derived above. The temperatures of the gas seen in the rovibrational lines are clearly higher than predicted from a similar extension of the temperature law, suggesting other sources of heating. Comparison of the radio and infrared data indicate that the line of sight to the infrared source has an unusually low column density.

[1]Electronic mail: carr@payne.mps.ohio-state.edu

[2]Visiting Astronomer at the Infrared Telescope Facility, which is operated by the University of Hawaii under contract from the National Aeronautics and Space Administration.

[3]Electronic mail: nje@astro.as.utexas.edu

[4]Electronic mail: lacy@astro.as.utexas.edu

[5]Electronic mail: zhou@sirius.astro.uiuc.edu

## 1. INTRODUCTION

Infrared absorption spectroscopy of interstellar molecules can provide some unique information about the chemical abundances and physical conditions in molecular clouds. In a previous paper (Evans, Lacy & Carr 1991, hereafter Paper I) we reported on mid-infrared molecular spectroscopy of $C_2H_2$, HCN, OCS, CO and $NH_3$ towards IRc2 and IRc7 in Orion. The absorption lines were found to form in gas with temperatures of 140-150 K and densities of $3 \times 10^6$ to $1 \times 10^7$ cm$^{-3}$ in either the plateau or hot core features. The abundances determined for $C_2H_2$ and HCN were about $10^{-7}$ to $10^{-6}$ of $H_2$, much larger than predictions of steady-state gas-phase chemical models but similar to predicted peak abundances obtained at early times, $t \simeq 1 - 3 \times 10^5$ yr. These results were consistent with models in which molecules with early-time abundances were frozen onto dust grains and later released into the gas after the dust was heated as a result of star formation.

Infrared spectroscopy provides a number of advantages for the study of molecular clouds. Molecules without permanent dipole moments which cannot be observed using radio spectroscopy, such as $C_2H_2$ and $CH_4$, can be studied via their rovibrational lines (Lacy et al. 1989a; Lacy et al. 1991). Several different rotational transitions of a molecule can be observed with relative ease, allowing the derivation of excitation temperature and column density without the problem of correcting for different beam sizes encountered in radio observations. In addition, infrared absorption measurements provide high spatial resolution, though one is limited to a line-of-sight towards an embedded or background infrared source.

In this paper, we apply these techniques to the study of $C_2H_2$, HCN, $NH_3$, CS, SO, and SiO towards the embedded young stellar object GL 2591. In addition, we combine the infrared data with radio-wavelength observations of HCN, CS, $H_2CO$ and $HCO^+$. GL 2591 is a luminous young stellar object obscured by about 70 magnitudes of visual extinction. Its luminosity is about $2 \times 10^4 L_\odot$ for a distance of 1 kpc (Mozurkewich et al. 1986). GL 2591 is associated with an extended high-velocity CO outflow (Bally & Lada 1983; Mitchell et al. 1991) and extremely high-velocity gas with a full-width of 75 km s$^{-1}$ in the CO line wings (Choi et al. 1993). High-velocity gas is also observed in absorption in the fundamental lines of CO (Geballe & Wade 1985; Mitchell et al. 1989) with blueshifted absorption extending to at least 100 km s$^{-1}$ in the $^{12}$CO $v = 0 - 1$ lines. Still higher velocities (500 km s$^{-1}$) are seen in spectra of Herbig-Haro objects (Poetzel, Mundt, & Ray 1992). The extensive work of Mitchell et al. (1989) identified a number of discrete CO absorption components and derived temperatures and column densities from $^{13}$CO $v = 0 - 1$ lines. The gas temperatures range from 38 K to 1000 K with a total CO column density of about $10^{19}$ cm$^{-2}$.

## 2. OBSERVATIONS AND RESULTS

### 2.1. Infrared

Observations were made on the NASA Infrared Telescope Facility in 1989 July, 1989 October, 1990 July, and 1991 November using the mid-infrared echelle spectrograph Irshell (Lacy et al. 1989b). Irshell uses a 64x10 detector array with 64 pixels in the spectral direction and 10 pixels along the slit with a scale of 1″ per pixel. The two spatial halves of the array are staggered by 0.5 pixels in the spectral direction. Hence, by positioning a point source in the slit such that it is centered between the two central rows of the array, a 128 point spectrum is acquired. Nearly all of the spectra of GL 2591 were taken in this mode. A slit width of 1″.5 was used for most of the observations, giving velocity resolutions of 26 to 32 km $s^{-1}$. The 1991 November observations used a setup with 2.4 times the dispersion per pixel but a similar velocity resolution. The data reduction procedure, described in Paper I, involved flat-fielding by a dome temperature blackbody followed by division by a comparison source spectrum to remove telluric absorption lines. The wavelength calibration was determined by fitting a linear wavelength scale to telluric absorption lines in the spectrum.

Spectra near 13 $\mu$m were obtained for R-branch and Q-branch lines of the $\nu_5$ band of $C_2H_2$ and for R-branch lines of the $\nu_2$ band of HCN. The spectra for most of the R-branch lines of $C_2H_2$ and HCN are presented in Figure 1. The $C_2H_2$ Q branch is shown in Figure 2. The measured equivalent widths and velocities for $C_2H_2$ and HCN lines are listed in Tables 1 and 2, respectively. Spectra were also obtained for lines of CS, SO, and SiO in the 8-9 $\mu$m region and lines of $NH_3$ near 10.5 $\mu$m. With the possible exception of the aR(0,0) line of $NH_3$ (see Fig. 1), none of these latter molecules were detected, and upper limits on the equivalent widths are given in Table 3. In all cases the uncertainties in measured equivalent widths are dominated by uncertainties in the placement of the continuum. Upper limits (2 $\sigma$) on non-detections were determined by the scatter in the continuum. The uncertainty in the radial velocities due to errors in the wavelength scale is typically $\pm$ 3 km $s^{-1}$, but larger systematic errors, such as correction for telluric lines and flat-fielding, probably dominate the measured velocities. Radial velocities are not listed for the higher dispersion measurements of the R(19) and R(21) lines of $C_2H_2$ since the smaller wavelength coverage did not provide enough telluric lines for wavelength calibration.

### 2.2. Radio

Observations were obtained with the 10.4-m telescope of the Caltech Submillimeter Observatory (CSO)[6] at Mauna Kea, Hawaii. The data were acquired on various runs between 1989, December and 1994, June. Maps were made of the $J = 3-2$ and $4-3$ lines of HCN with 21–26 positions; observations of several positions were obtained in the $H^{13}CN$ $J = 3-2$ line. A similar size map was made in the $HCO^+$ $J = 3-2$ line and a 13 point map was made in the $H^{13}CO^+$ $J = 4-3$ line. The $J_{K_{-1}K_1} = 5_{15} - 4_{14}$ (hereafter $J = 5-4$) line of $H_2CO$ was observed in the opposite sideband of the HCN $J = 4-3$ observations, and the $J_{K_{-1}K_1} = 3_{12} - 2_{11}$ (hereafter $J = 3-2$) line of $H_2CO$ was also observed at a single position. Maps with 11–17 positions were obtained in the $J = 5-4$ and $7-6$ lines of CS and observations at the peak position were obtained for CS $J = 10-9$ and $C^{34}S$ $J = 5-4$. In addition, we make use of data on the $J = 2-1$, $3-2$, and $5-4$ lines of CS and the $J = 2-1$ and $3-2$ lines of $C^{34}S$ observed with the IRAM 30-m telescope (Plume et al. 1995) and CO $J = 2-1$ and $3-2$ data from Choi et al. (1993).

The data were calibrated to the $T_A^*$ scale using chopper-wheel calibration. The $J = 10-9$ line of CS was corrected for different atmospheric opacities in the two receiver sidebands. For the IRAM data, the observations were scaled to achieve consistency with line calibration sources and further divided by the main beam efficiency ($\eta_{mb}$). Thus, IRAM data are shown as $T_{mb}$, while all other data are shown as $T_A^*$. There is an unusually large calibration uncertainty in the $J = 3-2$ line (Plume et al. 1995), and we tend to discount this line in our modeling. The $J = 2-1$ CS line may also be weaker than indicated by the IRAM data, since Yamashita et al. (1987) got a value of $T_{mb}$ of 7.6 K, lower by 35% than the value obtained at IRAM, even though their beam was slightly smaller. Telescope and line parameters are given in Table 4. Maps of six of the lines are shown in Figure 3. All the maps peak on the position of the infrared source within reasonable pointing uncertainties. Maps with the highest signal-to-noise (but spatial resolution of about 26″) are very centrosymmetric. Maps with higher spatial resolution (about 20″) may show some east-west elongation, but they have lower signal-to-noise. In general, the source shows little structure and is sharply peaked on the infrared source. Consequently, we treated any offsets in maps as pointing errors and shifted data to align the peaks for subsequent analysis.

---

[6]The CSO is operated by the California Institute of Technology under funding from the National Science Foundation, contract AST 90–15755.

## 3. ANALYSIS OF INFRARED DATA

In this section we derive column densities for $C_2H_2$ and HCN and place upper limits on the column densities for the other observed molecules. The results are summarized in Table 5. A description of the analysis and the basic molecular data are given in Paper I. Information on molecules not analyzed in Paper I is given in Appendix A.

### 3.1. $C_2H_2$

#### *3.1.1. R-Branch Lines*

We measured seven R-branch lines of the $\nu_5$ band of $C_2H_2$ (Table 1). The R(19) and R(21) lines were observed twice, and the averages of these measurements were used in our analysis. First, the temperature and column density were derived under the assumption that the lines are optically thin. In Figure 4 the natural logarithm of the column density per degenerate sublevel ($N_J/g_Jg_I$) is plotted versus the energy of the lower state. A linear fit to the data gives a temperature of 410 ± 40 K and $N(C_2H_2) = (1.2 \pm 0.2) \times 10^{16}$ cm$^{-2}$. The reduced $\chi^2$ for this fit is $\chi^2_r = 1.07$ ( $\chi^2_r = \chi^2/(N_{dat} - \nu)$, where $N_{dat}$ is the number of data points and $\nu$ the number of parameters). There is no evidence for saturation in the lines as was found for the $C_2H_2$ lines in IRc2 (Paper I), but $N(C_2H_2)$ is about an order of magnitude smaller in GL 2591. Attempts to fit the GL 2591 data with unresolved narrow lines and high optical depths only degraded the quality of the fit.

In their study of the fundamental lines of CO toward GL2591, Mitchell et al. (1989) observed several absorption components of different temperature and velocity. The three main components have temperatures and $V_{LSR}$ of 38 K and −11 km s$^{-1}$, 200 K and −28 km s$^{-1}$, and 1010 K and −8 km s$^{-1}$. The column density is similar in each component. Mitchell et al. (1990) thought that the cold gas could represent the molecular cloud core since the temperature and column density are similar to those derived from radio measurements of CO, but the sizeable velocity difference between the −11 km s$^{-1}$ absorption feature and the central velocity of the millimeter emission was puzzling. The other CO absorption components must be heated gas near the central object. The temperature of 410 K derived for $C_2H_2$ does not match any of the CO absorption components. The mean velocity for the seven $C_2H_2$ lines is $V_{LSR} = -13.7$ km s$^{-1}$ with a standard deviation ($\sigma$) of 5.2 km s$^{-1}$. This velocity is consistent with the cold gas at $V_{LSR} = -11$ km s$^{-1}$ and marginally consistent with the 1010 K gas at $V_{LSR} = -8$ km s$^{-1}$. However, none of the individual CO components agree with $C_2H_2$ in both velocity and temperature.

For any reasonable abundance ratio of $C_2H_2$ to CO, the gas which produces the $C_2H_2$ absorption should have a corresponding component observable in CO. The observed $C_2H_2$ is probably due to a combination of different temperature components which correspond to the observed CO components. Accordingly, we attempted to fit the $C_2H_2$ data with a multiple temperature model. The temperatures were fixed to those determined from the CO observations (38, 200, and 1010 K), and the three column densities were varied to obtain the best fit to the data. Since the lines are optically thin, the observed equivalent width of a line will be proportional to the sum of the column densities ($N_J$) in the corresponding rotational level. The minimum $\chi^2$, $\chi^2_{min}$, was found for $N_{38} = 0$, $N_{200} = (4.2 \pm 0.7) \times 10^{15}$ cm$^{-2}$, and $N_{1010} = (1.05 \pm 0.16) \times 10^{16}$ cm$^{-2}$. The best two-temperature fit to the rotational levels is presented in Figure 5. The uncertainties in each parameter come from the 1 $\sigma$ confidence interval found by projecting the $\chi^2_{min} + 1$ ellipse in the $N_{200} - N_{1010}$ plane onto each axis. The goodness of fit, $\chi^2_r = 1.33$, is similar to the single temperature fit. Comparison of Figures 4 and 5 shows that the combination of 200 K and 1010 K gas mimics a single component at an intermediate temperature.

Are the radial velocities also consistent with this two-temperature fit? The 200 K and 1010 K CO components have $V_{LSR} = -28$ and $-8$ km s$^{-1}$, respectively. Our 30 km s$^{-1}$ resolution is not sufficient to resolve these components, but the presence of two velocity components should be reflected in the measured radial velocities of the lines. For each line, the component velocities were weighted by the $N_J$ of each component, as determined by the two-temperature model, to give the expected radial velocity of the blend. The predicted velocity varies with $J$, from $V_{LSR} = -21$ km s$^{-1}$ in the R(1) line to $-10$ km s$^{-1}$ in the R(21) line; the expected mean velocity of the seven lines is $V_{LSR} = -15.1$ km s$^{-1}$ with $\sigma = 4.5$ km s$^{-1}$. This is consistent with the observed mean velocity of $-13.7\pm$ 5.2 km s$^{-1}$. The uncertainty in the velocity measurement for a single line, however, is too large to establish whether the expected trend of velocity with $J$ is present.

An upper limit was placed on the column density in the 38 K gas by increasing $N_{38}$ from zero in the three-temperature fit, while varying $N_{200}$ and $N_{1010}$ to minimize $\chi^2$, until $\chi^2$ was equal to $\chi^2_{min} + 4$ (corresponding to 2 $\sigma$ ). This gave a 2 $\sigma$ limit on $N_{38}$ of $6.0 \times 10^{14}$ cm$^{-2}$, with corresponding values of $N_{200} = 3.0 \times 10^{15}$ cm$^{-2}$ and $N_{1010} = 1.1 \times 10^{16}$ cm$^{-2}$. A three-temperature model with this combination of column densities is shown in Figure 5 by the dashed line. The 38 K gas produces significant population only in the lower rotational levels, and as a result the limit on $N_{38}$ depends largely on the equivalent width and the assigned uncertainty for the R(1) line. The R(1) point falls far below the fits in both Figures 4 and 5; it is 4 $\sigma$ below the best two temperature fit and nearly 7 $\sigma$ below the limit on $N_{38}$. Saturation in the lower rotational levels cannot account for this. Examination of the spectrum shows that the R(1) line falls on the steep shoulder of a telluric absorption line, which makes the result sensitive to the division by the standard; it is possible that the R(1) equivalent width is badly in error.

Confirmation of the population in the lowest J levels would be useful. We observed the R(0) line of the $\nu_4 + \nu_5$ band and placed a 2 $\sigma$ upper limit on $w_v$ of $1.3 \times 10^{-3}$ cm$^{-1}$. However, the transition probabilities for the $\nu_4 + \nu_5$ band are far smaller than those for the $\nu_5$ band, and the R(0) line does not place an interesting constraint on the column density of cold $C_2H_2$: if all of the J = 0 gas is assumed to be at 38 K, then the 2 $\sigma$ limit on $N_{38} = 1.9 \times 10^{16}$ cm$^{-2}$. A better handle on the amount of cold $C_2H_2$ comes from analysis of the Q branch in the next section.

### *3.1.2. The Q Branch*

The observed Q branch of the $\nu_5$ band is shown in Figure 2. The Q branch lies in a region of poor atmospheric transmission (Fig. 2a), including a set of strong water lines overlapping with the lower J lines of the band. The observations were made possible by reasonably dry conditions, about 1 mm of precipitable water vapor. The useful region of the Q branch covered by our spectrum includes the Q(1) through the Q(11) lines. The wavenumbers for the odd-J lines are marked in Figure 2b for $V_{LSR} = -14$ km s$^{-1}$, the mean velocity observed for the R-branch lines. Note that the even-J lines are not prominent in the data due to their lower nuclear statistical weights. The continuum level in the telluric-corrected spectrum was set to the mean level at the transmission peak redward of the Q branch at 729.05 cm$^{-1}$. Unfortunately, from the point of view of constraining the column density in the 38 K gas, it is the lowest J lines that are most affected by the telluric water lines.

Synthetic Q-branch spectra were calculated for comparison with the data by assuming the lines to be optically thin, which was justified by the $N_J$ found in the analysis of the R branch. For the date of observation, 1989 July 25, the correction between $V_{LSR}$ and observed velocity was $-23.4$ km s$^{-1}$. The synthetic spectra were smoothed with a velocity resolution of 26 km s$^{-1}$. Figure 2b compares the data to a model using the best single temperature fit to the R-branch lines, $T = 410$ K and $N = 1.2 \times 10^{16}$ cm$^{-2}$, where the mean $V_{LSR} = -14$ km s$^{-1}$ was used. Clearly, this provides a poor fit to the observed spectrum. For two and three-temperature models, the temperatures and their respective radial velocities were fixed at the values for the CO absorption lines. A synthetic spectrum using the best two-temperature fit to the R-branch lines is shown in Figure 2c. This spectrum also included the HCN R(5) line by using the two-component solution from §3.2. The synthetic spectrum based on the two-temperature model provides a better match to the shape of the Q branch spectrum and a rough confirmation of the column densities from the R-branch analysis. Recall that the one-temperature and two-temperature fits to the R-branch lines were nearly indistinguishable, but the latter is clearly preferred by the Q-branch data. This is partly due to the broader lines that result from the differing velocities of the two components. However, the model in Figure 2c does seem to require more column density in the lowest J levels. In Figure 2d, we show the three temperature model using the previously derived limit on the cold gas $N_{38} = 6.0 \times 10^{14}$ cm$^{-2}$, with $N_{200} = 3.0 \times 10^{15}$ cm$^{-2}$ and $N_{1010} = 1.1 \times 10^{16}$ cm$^{-2}$. This improves the match to the data. The dashed lines in Figure 2d show the result of increasing $N_{38}$ to $10 \times 10^{14}$ cm$^{-2}$ and to $30 \times 10^{14}$ cm$^{-2}$. $N_{38} = 10 \times 10^{14}$ cm$^{-2}$ is somewhat too large, and $N_{38} = 30 \times 10^{14}$ cm$^{-2}$, equal to $N_{200}$, is clearly ruled out.

The important result from the Q-branch analysis is the confirmation that the column density of $C_2H_2$ in the cold gas is significantly lower, by about an order of magnitude, than the average column density in the hot gas components. Since the CO column densities for each component are nearly identical, the abundances of $C_2H_2$ in the hot and cold gas are significantly different. The Q-branch data imply a column density in the cold gas that is consistent with the limit established by analysis of the R branch. Keeping in mind the poor atmospheric transmission at the wavelengths of the Q branch, we adopt $N_{38} = 8 \times 10^{14}$ cm$^{-2}$ as a tentative detection of cold $C_2H_2$. The column density in the 38 K gas could be better established by a measurement of the R(1) line at a time when GL 2591 has a more favorable velocity shift.

## 3.2. HCN

We observed 6 lines of the R branch of the $\nu_2$ band of HCN (Table 2). The initial observations provided very marginal detections of the R(21) and R(22) lines. A second spectrum of the R(21) line was obtained at a time of the year when the source velocity with respect to telluric lines was more favorable. This spectrum gave an upper limit on the R(21) equivalent width 4 times lower than the earlier measurement; this result also makes the $1.5\sigma$ detection of the R(22) line questionable. In the initial fitting of the HCN data to determine column densities, we used only the four clearly detected lower rotational lines. First, the data were fitted with a linear relationship assuming a single-temperature, optically thin component. This result is shown in Figure 6, with the best fit straight line giving T = 303 ± 38 K and $N$(HCN) = $(3.1 \pm 0.6) \times 10^{16}$ cm$^{-2}$, with $\chi_r^2 = 1.18$. As we found for $C_2H_2$, the derived temperature is intermediate between the 200 K and 1010 K temperatures of components seen in CO.

Next, we applied the same three-temperature analysis used for $C_2H_2$. Again, the minimum $\chi^2$ was found for zero column density in the 38 K gas. Figure 7 shows the best two-temperature fit, in which the column densities are $N_{200}$(HCN) = $(2.0 \pm 0.4) \times 10^{16}$ cm$^{-2}$ and $N_{1010}$(HCN) = $(1.6 \pm 0.5) \times 10^{16}$ cm$^{-2}$, with $\chi_r^2 = 0.63$. Nearly the same HCN column density is required for the 200 K and 1010 K gas, in contrast to $C_2H_2$, for which $N_{1010}(C_2H_2)$ was about 2.5 times $N_{200}(C_2H_2)$. A 2 $\sigma$ upper limit of $1.7 \times 10^{16}$ cm$^{-2}$ was placed on $N_{38}$(HCN) using the same procedure that was applied to $C_2H_2$. The dashed line in Figure 7 shows the limit for $N_{38}$(HCN) with the corresponding column densities $N_{200}$(HCN) = $1.3 \times 10^{16}$ cm$^{-2}$ and $N_{1010}$(HCN) = $2.1 \times 10^{16}$ cm$^{-2}$. Note that the limit on $N_{38}$(HCN) is comparable to the column densities measured in the hot gas. The reason for the poor constraint on $N_{38}$(HCN) is that the lowest rotational line observed, R(6), arises from a state 89 K above ground and hence is not very sensitive to the presence of cold gas.

The mean $V_{LSR}$ for the four measured HCN lines is $-8.6$ km s$^{-1}$ $\pm 5.0$ km s$^{-1}$. This velocity is consistent with either the 38 K or 1010 K CO velocities. For the column densities in the best two-temperature fit, the expected mean velocity of these same lines is $-19.8$ km s$^{-1}$. Unlike the good agreement that was found for $C_2H_2$, the observed velocities for HCN disagree with the two-temperature model by $2\sigma$. Higher dispersion spectra are required to test our interpretation of the data.

The above fits did not consider the R(21) and R(22) lines, but the best two-temperature fit predicts a population in the higher J lines significantly larger than the 2 $\sigma$ limit on the R(21) line. This departure could be due to subthermal population of the higher rotational levels. To investigate this, we modeled the population in the rotational levels using a Large-Velocity-Gradient (LVG) code with 29 levels. The HCN collisional rates were provided by Green (1994, private comm.). A velocity width (FWHM) of 12.9 km s$^{-1}$ was assumed, which is similar to the CO line widths measured by Mitchell et al. (1989) for the 200 and 1010 K gas. The temperatures of the components were fixed at 200 and 1010 K. This leaves two column densities and two volume densities ($n = n(H_2) + n(He) + \ldots$) as variables, which are too many free parameters to constrain uniquely with the data. We assume the same densities in the two gas components and consider what density is required to explain the departure of the high J lines. We have not carried out a systematic search of parameter space, but densities of $2 - 3 \times 10^7$ cm$^{-3}$ best match the data for the above choice of line width. Figure 8 illustrates three models with slightly different densities, in all of which $N_{200}$(HCN) is reduced somewhat from its best LTE value to $1.6 \times 10^{16}$ cm$^{-2}$, but $N_{1010}$(HCN) remains the same at $1.6 \times 10^{16}$ cm$^{-2}$. A value around $n = 3 \times 10^7$ cm$^{-3}$ probably gives the best fit. While these results are not uniquely constrained, they do show that the limit on the R(21) line is consistent with the existence of a gas component at 1010K when subthermal excitation is taken into account, that the LTE column densities for HCN are not greatly in error, and that the gas densities in the HCN-absorbing region are on the order of $3 \times 10^7$ cm$^{-3}$, consistent with the estimate that $n \geq 10^7$ cm$^{-3}$ based on $^{13}$CO (Mitchell et al. 1989).

### 3.3. Other Molecules

We made observations towards GL 2591 of molecular transitions of CS, $NH_3$, SO and SiO, but no clear detections were made. An upper limit on $CH_4$ was reported in a previous paper (Lacy et al. 1991). In Table 3 we list the molecules and transitions observed and the 2 $\sigma$ upper limits on their equivalent widths. Limits were placed on the column density in the gas by assuming the same temperatures for the CO components used in our analysis of $C_2H_2$ and HCN. These results are collected in Table 5.

The data for the R(0) line of the $\nu_4$ band of $CH_4$ in GL 2591 were reported in Lacy et al. (1991). In this paper, we use a slightly revised 2 $\sigma$ limit on the equivalent width of $8 \times 10^{-3}$ cm$^{-1}$. The column densities (Table 5) were calculated for temperatures of 38, 200, and 1010 K using the line strengths of Champion et al. (1989) in the manner described by Lacy et al. (1991). Since only the R(0) line was observed, the limits are most stringent for the cold gas and are not very interesting for the 1010 K gas.

Three wavelength settings were observed for $NH_3$, covering the aR(0,0) line, the sP(3,K) lines, and part of the sQ branch from J = 1 to 6. A feature was observed at the position of the aR(0,0) line (see Fig. 1) with a strength of $(1.5 \pm 0.5) \times 10^{-3}$ cm$^{-1}$ and $V_{LSR} = -9.5$ km s$^{-1}$; we regard this as an uncertain detection of the aR(0,0) line. No other $NH_3$ lines were detected, and a 2 $\sigma$ limit of $0.8 \times 10^{-3}$ cm$^{-1}$ was placed on their equivalent widths, excluding sQ-branch lines that fall within telluric absorption features. For each transition, the equivalent width was converted into a total column density $N(NH_3)$ for temperatures of 38, 200, and 1010 K as described in Appendix A. The aR(0,0) line, if real, must be due entirely to the 38 K gas, since higher energy transitions would be easily detected if the hot gas produced the observed equivalent width in the aR(0,0) line. The column density would then be $N_{38}(NH_3) = (5 \pm 2) \times 10^{14}$ cm$^{-2}$. Analysis of $NH_3$ emission at radio wavelengths gave a column density of at least $1.5 \times 10^{14}$ cm$^{-2}$ (Takano et al. 1986). The maximum rotational temperatures for the radio $NH_3$ lines are about 250 K (Torrelles et al. 1989), similar to the 200 K CO absorption component. The constraints on $N(NH_3)$ in Table 5 for the 200 and 1010 K gas were set using the transitions that were most favorable for each temperature.

Spectra were obtained which also included: the R(7) and R(11) lines of CS; the R(0), R(1), and P(15) lines of SiO; the $R_3(2)$ line and the R(1) and R(0) lines of SO. The upper limits on these lines are listed in Table 3, and the corresponding column density limits for assumed temperatures of 38, 200 and 1010 K are given in Table 5, except for SO, for which we lack a band strength.

## 4. ANALYSIS OF RADIO OBSERVATIONS

When appropriate, the data for each molecule were resampled to achieve a common spectral resolution: 0.83 km s$^{-1}$ for HCN and $H_2CO$ $J = 5-4$; 0.3 km s$^{-1}$ for $HCO^+$, $H_2CO$ $J = 3-2$, and the CS transitions other than the $J = 7-6$ (0.43 km s$^{-1}$) and $J = 10-9$ (0.70 km s$^{-1}$). The RMS noise in Table 4 refers to the noise in the baseline of the resampled spectra at the center position. Linear baselines were removed and Gaussians were fitted to each line. In some cases, data were also smoothed spatially to achieve comparable spatial resolution for the first step of the analysis.

### 4.1. Single-density Models

The first step was to try to fit the CS and HCN data with a model of a cloud with constant density, temperature, and column density. The temperature was fixed at 53 K, in order to match the CO observations of Choi et al. (1993) with comparable spatial resolution; the results of this analysis are not very sensitive to the exact choice of $T$. A grid of models was run with a LVG code, covering the density range of $n = 10^5$ to $10^8$ cm$^{-3}$ and the range $N/\Delta v = 10^{11}$ to $10^{15}$ cm$^{-2}$ (km s$^{-1}$)$^{-1}$ in column density per unit velocity interval, the quantity which is proportional to optical depth. The best-fitting values of $n$ and $N$ were determined by minimizing $\chi^2$. For HCN, with only two lines, the best fit was obtained for $n = 6 \times 10^6$ cm$^{-3}$, five times lower than the density obtained from the analysis of the infrared lines. However, once the $H^{13}CN$ data were included in the fitting process, $\chi^2$ degraded very badly, unless the isotope ratio was unreasonably small (HCN/$H^{13}CN$< 15). The situation for CS was even worse, even when only those data with similar spatial resolution were included (the $J = 2-1$ from IRAM, and the $J = 5-4$ and $7-6$ lines from the CSO). The best fit with the CS data only was similar to that for the HCN ($n = 4 \times 10^6$ cm$^{-3}$), but $\chi^2$ was high. When the $C^{34}S$ data were included, the fit became still worse. The problem was that high density solutions produced too weak a $J = 2-1$ line, while lower density solutions produced too weak a $J = 7-6$ line. These symptoms all pointed to the need for density (and perhaps temperature) inhomogeneities.

## 4.2. Models with Density Gradients

While unresolved clumps could be present, a model with smooth temperature and density gradients is also plausible in regions of star formation. Certainly, one expects the temperature to fall off with distance from the infrared source. In addition, there is observational evidence for density gradients. The densities derived from the radio data are less than those found from the analysis of the infrared absorption, which presumably comes from a smaller region. Furthermore, the $J = 5-4$ line from IRAM (observed with a smaller beam) is considerably stronger than the same line observed with the larger beam of the CSO. These facts justified an attempt to match the data with a cloud model with density and temperature gradients. To model clouds with gradients, we used a newly-developed one-dimensional Monte Carlo code (Choi et al. 1995), which allows arbitrary radial distributions of temperature, density, abundance, and velocity field. The centrosymmetric appearance of the maps (Fig. 3) supported the use of a one-dimensional geometry, at least on the scales probed by the radio emission. Collision rates were taken from Turner et al. (1992) for CS, Green and Thaddeus (1974) for HCN, Monteiro (1985) for $HCO^+$, and Green (1988) for $H_2CO$.

The models were constrained in a variety of ways. First, only power-laws in temperature and density were modeled, and the abundance of the molecule being modeled ($X = n(mol)/n$) was not allowed to vary with radius. The gas kinetic temperature ($T$) was assumed equal to the dust temperature, which was in turn assumed to behave as follows: $T_D(r) = 13\mathrm{K} r_{pc}^{-0.4}$. The exponent is expected for optically thin emission from grains with opacity decreasing linearly with wavelength. The optically thin approximation is reasonable for the range of temperatures considered here. The coefficient was determined by scaling the $T_D(r)$ curve computed for NGC2071 using a radiative transport code (Butner et al. 1990), by $(L(GL2591)/L(NGC2071))^{0.2}$ (see, e.g., Doty & Leung 1994). For all of this modeling, a source distance of 1 kpc and a luminosity of $2 \times 10^4$ $L_\odot$ (Mozurkewich et al. 1986) was assumed. The assumed temperature gradient did produce temperatures consistent with the observed CO $J = 2-1$ and $3-2$ lines (Choi et al. 1993).

The outer radius of the model was set at 0.3 pc ($1'$), sufficient to model our maps, the largest of which had a radius of $45''$. The model included 40 radial shells, equally spaced by $7.5 \times 10^{-3}$ pc ($1''.5$). Values of density, temperature, etc. are specified at the mean, mass-weighted radius of each shell, avoiding any problem with singularities at the origin. The temperature ranged from 100 K to 21 K. For the innermost shells, the actual temperature might exceed that from the optically thin approximation, but a full dust radiative transport calculation would be needed to calculate it. Since the inner regions are not very important to this modeling and since other sources of heating are likely present, based on the infrared data, we stayed with the simple power law.

For simplicity, the models used a purely microturbulent velocity field, and the width was adjusted to give a reasonable match to the CS data, resulting in a $e^{-1}$ width of 1.6 km s$^{-1}$. Models with density power laws ($n(r) = n_1 r_{pc}^{-\alpha}$) were used, with $\alpha$ of 1.0, 1.5, and 2.0 ($n_1$ is the density normalized at 1 pc). Within a set of models for a given $\alpha$, the free parameters were $n_1$ and $X$, the abundance of the molecule. Since we were also using isotope data to constrain the fit, we fixed the isotope ratio $R$ at 20 for CS/C$^{34}$S and 55 for HCN/H$^{13}$CN and HCO$^+$/H$^{13}$CO$^+$, the latter based on studies of the C/$^{13}$C ratio as a function of galactocentric distance (Langer & Penzias 1990) and the assumed galactocentric distance (8.4 kpc) of GL2591 (see Choi et al. 1993).

We fitted the CS data, which provided more constraints, first and then tested whether the same $n_1$ could fit the HCN data by varying only $X$(HCN). Finally, we modeled the HCO$^+$ and $H_2$CO data by varying only their abundances. Models with $\alpha = 1.5$ worked best overall. We will present and discuss those first, followed by a discussion of models with $\alpha$ =1.0 and 2.0. For any given $\alpha$, we ran models for a grid of $n_1$ and $X$. The best fit was determined by comparing the model line profiles to the observations by eye, taking into account the calibration uncertainties in the different lines. The best-fitting model in the $\alpha = 1.5$ category is one with $n_1 = 1 \times 10^4$ cm$^{-3}$ and $X(\mathrm{CS}) = 4 \times 10^{-10}$. The CS line profiles from this fit are compared to the observed profiles in Figure 9. The most discrepant line is the $J = 2 - 1$ line observed at IRAM, but the calibration uncertainty in this line is sufficient to explain this discrepancy. The model can produce the general pattern of line strengths and also the very different line strengths of the $J = 5 - 4$ line, when observed with different resolution. This test is one which distinguishes models with different $\alpha$. Even the slightly narrower C$^{34}$S lines, compared to the CS lines, come out of the microturbulent model, but the profiles of the lower $J$ lines of CS are starting to show some flattening, characteristic of modest optical depths in microturbulent models. It is hard to assign uncertainties to the best-fitting parameters, but variations by factors of 2 give clearly worse fits.

The HCN lines are shown in Figure 10. The line strengths are roughly correct and the wider lines in the HCN main isotope are even roughly reproduced, but the line profiles in the main line are very strongly self-reversed, an effect not seen in the data. This discrepancy is a well-known problem for purely microturbulent models of very opaque lines. We discuss this problem later. Meanwhile, we note that the HCN abundance is mostly controlled by the H$^{13}$CN line and the assumed isotope ratio of 55. If the actual isotope ratio in HCN is less than 55, $X$(HCN) would be lower, but the abundance is unlikely to be much higher. The main conclusions are that the same density law which fits the CS can fit the HCN with an HCN abundance of $X(\mathrm{HCN}) \sim 7 \times 10^{-10}$.

Also shown in Figure 10 are lines of $H_2$CO and HCO$^+$. The $H_2$CO lines are both well-fitted by the density law which fits the CS, with $X(\mathrm{H_2CO}) = 2 \times 10^{-10}$. The HCO$^+$ lines are less satisfactorily fitted. The best compromise is with $X(\mathrm{HCO^+}) = 4 \times 10^{-10}$, but the model profile at the center is too weak and self-reversed. As with HCN, the abundance is mostly controlled by the optically thin H$^{13}$CO$^+$ line and the assumed isotopic ratio.

Line profiles were also computed for positions displaced from the center of the map and compared to spectra obtained by averaging all the observed spectra at the same offset. This test was done for CS $J = 5 - 4$ and $7 - 6$, HCN $J = 3 - 2$ and $4 - 3$, and HCO$^+$. On this test, the $\alpha = 1.5$ model reproduced the off-center HCN lines and the CS $J = 5 - 4$ line quite well but failed by a factor of two to reproduce the $J = 7 - 6$ CS line. In contrast, the $HCO^+$ line at a position offset by $36''$ is too strong by a factor of about two. The $HCO^+$ data might be better fitted by a steeper density law or a varying abundance of $HCO^+$.

Monte Carlo techniques necessarily have uncertainties associated with the random number seed (Choi et al. 1995). To assess the uncertainties, models were run with the best-fit input parameters for ten different choices of random number seed. The average standard deviation of the populations was 1% of the population. Use of the average of the ten simulations in calculating model line profiles, rather than any one of the ten, produced no measurable differences. This source of uncertainty is negligible compared to those discussed below.

### 4.3. Alternative Models

In this subsection, we consider models with different choices for $\alpha$, the problem of self-reversed lines, and models with central holes. We also compare the column densities in the models to those inferred from continuum studies.

The best-fitting model in the $\alpha = 1.0$ category had $n_1 = 7.5 \times 10^4$ cm$^{-3}$ and $X(CS) = 2 \times 10^{-10}$. The line profiles match the observed profiles reasonably well (Fig. 11) with the exception of the IRAM $J = 3 - 2$ data, which we discount, as explained above. The other noteworthy discrepancy is that the ratio of the $J = 5 - 4$ lines in the two different beams proves harder to match with $\alpha = 1$. The model line is too strong in the big CSO beam and too weak in the small IRAM beam. Keeping $n_1$ fixed, models were run for HCN, varying only $X$(HCN). The best fit was obtained for $X$(HCN) $= 3.0 \times 10^{-10}$, with about a factor of 3 uncertainty (Fig. 12). The comments about self-reversed lines and isotope ratios made for $\alpha = 1.5$ apply equally in this case. The $H_2CO$ lines are fitted less well than they were by the $\alpha = 1.5$ model, and the $HCO^+$ is again not very satisfactory. The $\alpha = 1$ model came closer to the observed off-center $J = 7 - 6$ line, but produced a CS $J = 5 - 4$ line that was about 30% too strong; off-center HCN lines were also noticeably too strong in the $\alpha = 1$ models. The predicted $HCO^+$ spectrum at the position displaced by $36''$ is much too strong. This problem, already present in the $\alpha = 1.5$ model, is worse in the $\alpha = 1$ model. Off-center spectra of $H^{13}CO^+$ (not shown) are matched better by $\alpha = 1.5$ models than by $\alpha = 1$ models.

The best-fitting model in the $\alpha = 2.0$ category had $n_1 = 1.0 \times 10^3$ cm$^{-3}$ and $X(CS) = 8 \times 10^{-10}$. The discrepancies between the model and the data (Fig. 13) are the opposite of those seen with the $\alpha = 1.0$ models. The model lines are too weak for data observed with larger beams (e.g., $J = 5 - 4$ data from the CSO and $J = 2 - 1$ data from IRAM) and too strong for the $J = 10 - 9$ data, observed with a small beam. The model lines are too weak in the off-center positions for all the lines (Fig. 14), including the $HCO^+$, which was too strong in the off-center position in the $\alpha = 1.5$ model. $HCO^+$ would probably be better matched by a value of $\alpha$ between 1.5 and 2.

All choices for $\alpha$ produced self-reversed lines for the main isotopes of HCN and $HCO^+$ for the assumed isotope ratio of 55 and an abundance of the $^{13}$C isotopomers which reproduced the observations. Lowering the isotope ratio ameliorates the self-reversal but makes some of the model lines too weak. The problem is that the lower excitation outer parts of the cloud block the emission from the higher excitation inner parts in a homogeneous cloud of constant abundance with no systematic velocity, a well-known problem (see, e.g., Goldreich and Kwan 1974). Solutions to this problem include abundance gradients, systematic velocity gradients, and changes in the cloud geometry or structure. Abundance gradients are likely on theoretical grounds, as the outer part of the cloud will be affected by photodissociation, but modeling these would introduce more free parameters. As a simple experiment, we decreased the abundance of HCN by a factor of 10 outside a radius of 45″, but this change had little effect on the profiles, showing that the self-reversal in the observed lines arises closer to the center. Since the preferred value for $\alpha$ of 1.5 is that expected for collapse, a velocity gradient is plausible. Introduction of a velocity gradient ($v(r) = -0.25$km s$^{-1}r^{-0.5}$) characteristic of collapse produces less self-reversed lines with peak values similar to the observations, but at blue-shifted velocities with respect to optically thin lines (see Zhou 1992). Since the blue-shift characteristic of collapse is not apparent in the observations, this solution does not seem satisfactory either. Deviations from strict spherical symmetry and homogeneity are probably involved in avoiding the self-reversed profiles, but modeling these requires a code capable of modeling clouds in two or three dimensions.

One geometrical variation we could explore was the introduction of a central hole. Observations of the near-infrared continuum show a loop extending to the east of the source (Forrest & Shure 1986) which might be a cavity (Tamura & Yamashita 1992). The high-resolution $NH_3$ $(J,K) = (1,1)$ map of Torrelles et al. (1989) also suggests a cavity of about 5″ radius, although the (2,2) data do not show a cavity, suggesting that the apparent cavity in the (1,1) data may be an effect of elevated temperature. We ran some models with abundances decreased by factors of 100 inside a radius of 0.03 pc (6″). The HCN lines were less self-reversed, but also too weak. The central hole made the lines from high-J levels of CS ($J = 7-6$ and $J = 10-9$) too weak, even with $\alpha = 2$. At any rate, the optimum abundances in models with holes differ by about a factor of 2 from our standard models.

The total column densities, averaged over the model cloud, can be computed by dividing the molecular column density by the abundance. For the $\alpha = 1.5$ models, the result is $N = 1.4 \times 10^{23}$ cm$^{-2}$, similar to that derived from dust continuum emission in a region about half the radius of our model (Walker, Adams & Lada 1990). For the $\alpha = 1.0$ models, $N = 4.7 \times 10^{23}$ cm$^{-2}$, about four times that derived from the dust continuum emission. The $\alpha = 2.0$ models have $N = 4 \times 10^{22}$ cm$^{-2}$, about three times less than derived from the dust continuum emission. These comparisons would tend to favor the $\alpha = 1.5$ models, but discrepancies between molecular and dust column densities may be caused by density inhomogeneities. All the models predict very large column densities ($N = 1 \times 10^{24}$ cm$^{-2}$) along the line of sight to the center of the cloud. The implied extinction ($A_V = 1000$ mag) is much greater than the extinction ($A_V = 69$) to the infrared source, estimated from the silicate feature (Willner et al. 1982) and the extinction law of Rieke & Lebofsky (1985). While coatings of $H_2O$ ice can wash out the silicate feature (Pollack et al. 1994), resulting in an underestimate of $A_V$, the extinction clearly must be much less than $A_V = 1000$ in order for us to observe the source at 8-13 $\mu$m. Models with central holes large enough to give $A_V = 69$ cannot produce the lines observed with small beams, indicating that the line of sight to the source must have unusually low extinction. Deviations from spherical symmetry or density inhomogeneities are likely explanations; these are also likely to play a role in eliminating the self-reversed profiles in some lines. One deviation from spherical symmetry that is known to be present is a bipolar outflow. The high radial velocities seen in the outflow indicate that the outflow must be partly along the line of sight. The resulting cavity would then lower $A_V$.

### 4.4. Conclusions

We conclude that $\alpha = 1.5$ models can do a reasonable job of reproducing the radio data. Models with $\alpha = 1.0$ or 2.0 do somewhat worse. For our purposes (calculating abundances in the extended envelope), the differences between choices of $\alpha$ are factors of two, similar to the uncertainties associated with optimizing the choice of $n_1$ and the presence or absence of a central hole. For HCN and $HCO^+$, an additional uncertainty arises from the isotope ratio, again about a factor of two. All models of a spherical, homogeneous cloud predict too much extinction to the infrared source, suggesting that deviations from spherical symmetry or density inhomogeneities are present. Inhomogeneities could cause uncertainties in the abundances larger than a factor of two. Abundances based on optically thin lines of rare isotopes (HCN, $HCO^+$, and, to some extent, CS) are less affected by density inhomogeneities.

## 5. DISCUSSION

### 5.1. Abundances

In order to discuss the abundances towards GL 2591 as derived from the infrared data, we will express abundances as a fraction of the CO column density. The CO column density is determined using absorption spectroscopy in the same manner as the column densities measured in this paper, and this allows a straightforward comparison for each component of the gas. In addition, estimates of the $H_2$ column densities are largely based on measures of the dust which may not be appropriate for the hot gas. The CO column densities from Mitchell et al. (1989) are given in Table 5. The abundances with respect to gaseous CO are given in Tables 6 and 7 for the cold gas and hot gas components, respectively.

For the hot gas (Table 7), the average abundance of the two components are used for $C_2H_2$ and HCN, while the tighter limits for the 200 K gas are used for $CH_4$ and $NH_3$. Included in Table 6 are measurements for solid $H_2O$ and CO. A total $H_2O$ ice column density of $1.7 \times 10^{18}$ cm$^{-2}$ was derived from the measured ice feature at 3.08 $\mu$m (Smith, Sellgren & Tokunaga 1989) and the optical constants in Hudgins et al. (1993). The ratio of solid $H_2O$ to total H column density in GL 2591 is similar to that observed towards other embedded YSOs (Tielens et al. 1991). Solid CO is not detected towards GL 2591, with a limit on the ratio of solid to gaseous CO of 0.004 (Mitchell et al. 1990). This limit is at the lower end of ratios measured towards other embedded YSOs (Tielens et al. 1991). Since grain mantles should be evaporated in regions of high temperature, the solid features are assumed to correspond to the cold CO gas.

The abundance of CO with respect to $H_2$ can be estimated for the cold gas. Using $A_V = 69$ (§4.3) and $N(H_2) = 10^{21} A_V$, gives $N(H_2) = 6.9 \times 10^{22}$. This estimate of $N(H_2)$ gives $X(CO) = 1 \times 10^{-4}$, consistent with the values from most chemical models for dense molecular clouds (see Table 6).

### 5.2. Comparison of Radio and Infrared Data

In this section, we compare the results of the radio and infrared analyses. First, we compare the absorption velocity of the cold ($T = 38$K) component seen in CO absorption to the radio emission profiles; the $-11$ km s$^{-1}$ velocity of the absorbing gas lies well outside the central profile of the radio lines, so it cannot represent the ambient cloud gas. Some of the radio line profiles show a blue-shifted wing which extends to about $-16$ km s$^{-1}$. Much more extensive wings are seen in CO (e.g., Choi et al. 1993). Presumably, the $-11$ km s$^{-1}$ feature is a part of this outflowing material. The absence of CO absorption at $-5.8$ km s$^{-1}$, the peak of the emission lines, has been a puzzle. The total column densities in front of the central source inferred from the models in §4 would imply $N(\mathrm{CO}) = 10^{20}$ cm$^{-2}$ if $X(\mathrm{CO}) = 10^{-4}$, far in excess of those actually seen in absorption at other velocities. However, the estimate of extinction discussed in §4.3 shows that the line of sight to the infrared source has unusually low column density. Scaling the predicted $N$(CO) by 69/1000 (observed/predicted $A_V$) gives $N(\mathrm{CO}) = 7 \times 10^{18}$ cm$^{-2}$, comparable to that seen in the cold gas at $-11$ km s$^{-1}$. Some depletion of CO in the ambient cloud (see §5.3) would decrease the expected column density.

Second, we can use the radio data to set an upper limit to the size of the region which is seen in absorption in the infrared, using the HCN molecule, for which we have data of both kinds. The HCN emission is reasonably well accounted for by the extended cloud around the infrared source. Even without accounting for the lower extinction to the central source, the column density of HCN through the center of the model cloud ($1.4 \times 10^{15}$ cm$^{-2}$) is at least an order of magnitude less than those derived from the infrared lines. Thus it is not surprising that the velocity of absorption differs from the velocity of emission in the case of HCN. On the other hand, the conditions required to match the populations seen in absorption would produce very strong radio emission lines if they were resolved. Thus we set an upper limit on the size of the infrared absorbing region from the beam dilution needed to make the predicted emission equal to what we observe. The resulting limits are $3''$ for the $T = 200$K component and $2''$ for the $T = 1010$K component. At the assumed distance of 1 kpc, these upper limits correspond to 2000–3000 AU.

Third, we can compare the densities. The values of $n_1$ given in §4.2 are the values normalized to $r = 1$ pc, so they do not represent the densities in the region being modeled ($7.5 \times 10^{-3}$ to 0.3 pc). ¿From the inner to the outer radius, the densities are $1.5 \times 10^7$ cm$^{-3}$ to $6.1 \times 10^4$ cm$^{-3}$ for the $\alpha = 1.5$ model. We can also extrapolate the density laws inward, asking for the radius at which the density equals that favored by the non-LTE analysis of the HCN populations, $n = 3 \times 10^7$ cm$^{-3}$. The corresponding radius is 1000 AU for the $\alpha = 1.5$ model. A similar analysis for the temperature yields radii of 220 and 4 AU for the 200 and 1010 K components respectively. These are unrealistic, especially for the 1010 K component, since $T_D(r)$ is expected to deviate from a power law at small radii (see Butner et al. 1990) and other heating mechanisms for the gas, such as shocks, are suggested by the velocity shift of the absorption lines and the presence of a strong outflow. The shocks may well have affected the densities as well, but the fact that the sizes at which the model densities match those inferred for the HCN-absorbing gas are only factors of 2–3 smaller than the upper limits suggests that the amount of compression need not be large.

Fourth, we compare abundances from the radio and infrared analyses of HCN. The radio data favor $X(\mathrm{HCN}) = 7 \times 10^{-10}$ to within a factor of 2; to compare to the infrared data, we assume $X(\mathrm{CO}) = 1 \times 10^{-4}$ (see §5.1). Thus, HCN/CO= $7 \times 10^{-6}$, a factor of 400 less than the ratio in the hot, infrared-absorbing gas. Clearly something dramatic has happened to the chemical abundances on small scales.

### 5.3. Comparison to Chemical Models

In this section, we compare the abundances derived from the radio and infrared observations to those predicted by chemical models. Several kinds of chemical models will be encountered. First, there are models with only gas-phase chemistry and no depletion onto grains, such as Herbst & Leung (1989) and Langer & Graedel (1989). Second, there are models with gas-phase chemistry and depletion onto grains (models by Hasegawa & Herbst 1993 without desorption). Third, there are models with gas-phase chemistry, depletion onto grains, and a sudden release when the grains are heated. Since the grains play only a passive role in these models, we use models without depletion calculated at early times ($t = 10^5$ y) from Herbst & Leung (1989), when the abundances of some species of interest are near their peak. Fourth, there are models with gas-phase chemistry, depletion onto grains, reactions on grains, and release. The release is steady, due to cosmic ray desorption in some models by Hasegawa & Herbst (1993), or sudden, due to grain heating (Brown et al. 1988; Brown 1990). Finally, there is the hot core model of Charnley, Tielens & Millar (1992) which follows the chemistry of the hot gas after grain mantle evaporation.

We first compare the abundances from the radio data with chemical models (see Table 8). The abundances from the radio data are all considerably less than the predictions of steady-state gas-phase models (Herbst & Leung 1989; Langer & Graedel 1989), implying that some depletion has occured in the ambient cloud. In the models by Hasegawa & Herbst (1993, hereafter HH), run for $n = 10^4$ cm$^{-3}$, depletion onto grains is considered; these models produce abundances higher than seen with the radio data for times less than $10^6$ y, but the model abundances drop for larger times, as molecules deplete. All models for $t = 10^7$ y predict abundances much lower than observed, even if cosmic ray desorption is included. In addition, at $10^7$ y CO is severely depleted, $X(\mathrm{CO}) \sim 10^{-9}$, compared to estimates of $\sim 10^{-4}$ for the observed value. These comparisons suggest that considerable depletion has occured during the cloud lifetime. If most of that lifetime has been spent at a density of $10^4$ cm$^{-3}$, the cloud around GL2591 has a chemical age of a few million years.

Next, we examine whether the measured abundances from the infrared are consistent with the chemical models. The abundances of the cold and hot phases are compared to various models in Tables 6 and 7. First consider the pure gas-phase models. The most complete set of models for comparision with the infrared data are the results for cold dense clouds of Herbst & Leung (1989). For the cold gas we have only limits (HCN, $CH_4$) or tentative detections ($C_2H_2$, $NH_3$). It can be seen from Table 6 that the $C_2H_2$ and $CH_4$ abundances are consistent with steady state abundances but are substantially smaller than the models at $10^5$ y (early time). The upper limit on HCN is consistent with either $10^5$ y or steady-state abundances, and the value for $NH_3$ is about a factor of 5 lower than either.

In the hot gas, however, the abundances of $C_2H_2$ and HCN are far larger than the steady-state abundances (Table 7); this result is true for other steady-state models as well, such as Langer and Graedel (1989). The hot $C_2H_2$ and HCN are more consistent with the early-time abundances, though the prediction for $C_2H_2$ is somewhat too high. The limits on $CH_4$ and $NH_3$ cannot distinguish between early-time and steady-state models. Note that a comparison between our abundances in the hot gas and the low temperature gas-phase models is valid only if we are observing gas that has been recently heated and has not been altered by high-temperature chemistry. For the Herbst and Leung (1989) model in Table 7 we have taken the ratio of the peak abundances at $10^5$ y to the somewhat larger steady-state CO abundance; hence the $10^5$ y abundances listed in Table 7 are about a factor of two lower than those listed in Table 6. This might be more appropriate if the molecules at earlier times are frozen onto grains and returned to the gas phase at a later time but the CO remains in the gas phase long enough to reach a steady-state abundance.

Hence, in GL 2591 the upper limits on the abundances measured in the infrared for the cold molecular gas are consistent with steady-state abundances, but the abundances in the hot molecular gas are closer to abundances at early times ($\sim 10^5$ y). This could be explained if the high abundances predicted at early times for some molecules were preserved by accretion onto grains and then restored to the gas after the evaporation of grain mantles following star formation. The cold gas should be more representative of the ambient molecular cloud prior to star formation, with abundances appropriate to steady-state models, but the radio data indicate depletion of the gas-phase abundances with respect to steady-state models. The infrared data for $NH_3$ also indicates some gas-phase depletion in the cold gas, though $C_2H_2$ appears to be consistent with the steady-state predictions without depletion.

A scenario in which the grains play a purely passive role in storing molecules may be sufficient to explain the gas-phase abundances; however, an active role for the grains is indicated by the amount of $H_2O$ ice observed. If gas with the peak $H_2O$ abundance predicted by gas-phase models was accreted onto the grains, it would be one order of magnitude short in accounting for the $H_2O$ ice (compare col. 2 and col. 3 of Table 6). GL 2591 is not unusual in this regard, since grain surface reactions are required to explain the strength of ice features observed towards other sources (d'Hendecourt, Allamandola & Greenberg 1985; Tielens & Hagen 1982).

Next we compare the data to selected models which incorporate both grain-surface reactions and gas-phase chemistry. The gas-grain chemical models of HH, with initial molecular hydrogen and cosmic-ray-induced desorption, are given in columns 5 and 6 of Tables 6 and 7 for times of $10^5$ and $10^6$ y. In column 7 is Case 3 from Brown (1990), in which the initial fraction of atomic hydrogen is $1 \times 10^{-4}$. Since only the final mantle composition is tabulated in Brown (1990), for comparison with the hot gas abundances (Table 7) we assume all molecules are released back into the gas phase, and for the cold gas we only compare the prediction for $H_2O$ ice (Table 6). The hot core chemistry model (model A) of Brown, Charnley & Millar (1988) is shown in column 8. In this model the cloud evolves in density from $3 \times 10^3$ to $1 \times 10^7$ cm$^{-3}$ over $10^6$ y while allowing accretion by the grains, after which the mantles are assumed to be released back into the gas phase; only a limited surface chemistry was used.

Considering first the cold gas (Table 6), the abundances in the gas-grain models of HH are larger than the observed limits on $C_2H_2$, $CH_4$, and $NH_3$ by factors of 3 to 30. The abundances from Brown et al. (column 8) are taken at the time of the peak gas-phase abundance prior to grain evaporation, about $4 \times 10^5$ y. These disagree with both the data and with the results of HH, but Brown et al. used a different density and grain chemistry. All of the gas-grain models produce more $H_2O$ ice than observed, though the model of HH at $10^5$ y is only a factor of three too high. In addition, the predicted ratio of CO ice to gas is 10 to 100 times too large. Note that the abundance of $H_2O$ ice towards GL 2591 is similar to other embedded young stellar objects (see Tielens et al. 1991), and the over-production of water ice appears to be a problem for most of the grain chemistry models.

For the hot gas (Table 7), we compare the data to the total gas+surface abundances in the HH models at $10^5$ y (column 5) and $10^6$ y (column 6) under the assumption that the molecules in the mantles have simply been returned to the gas. The predicted abundances at $10^5$ y for $C_2H_2$, $CH_4$ and $NH_3$ are within factors of 5 or less of the observed values, similar to the pure gas-phase model at the same age, but large amounts of HCN are produced on the grains, in contradiction to the observations. The abundances apparently increase between $10^5$ and $10^6$ y, but this is only with respect to CO due to the conversion of CO on the grains to other species, and the abundances with respect to $H_2$ are nearly constant. Brown (1990) used a rather different surface chemistry compared with HH; not surprisingly, the final abundances differ, with less $CH_4$ and more $NH_3$ produced; both $C_2H_2$ and HCN are over-produced. The hot core model of Brown et al. (1988) produces far too much $CH_4$ and $NH_3$, but a very limited surface chemistry was used that only allowed hydrogenation of atoms accreted by the dust. Models where the hydrogen is initially atomic (HH model A(2100,CR); d'Hendecourt et al. 1985) also produce $CH_4$ and $NH_3$ with abundances comparable to CO; such models are clearly ruled out by the observed limits on $CH_4$ and $NH_3$, unless these molecules are rapidly destroyed after evaporation of the grain mantles.

While the observations are matched reasonably well by early-time low-temperature gas-phase chemistry with temporary storage in grain mantles, recent calculations by Herbst et al. (1994) examining the effect of rapid neutral-neutral reactions show that the peak abundances of many molecules could be greatly reduced. Compared to the previous results of Herbst & Leung (1989), the abundances in Herbst et al. (1994) for $C_2H_2$ and $NH_3$ are down a factor of 10 and the peak HCN abundance is decreased by two orders of magnitude. As regards the infrared results for GL 2591, only HCN is in severe disagreement. The measured $C_2H_2$ abundance is only a factor of four too large, and the disagreement is just a factor of two when the UMIST reaction network is used (Herbst et al. 1994). The C + $C_2H_2$ reaction appears to be important in reducing the abundance of acetylene and more complex hydrocarbons, but the rate coefficients for the neutral carbon reactions have only been measured at room temperature, and extrapolation to low temperatures is controversial. More recently, however, Bettens, Lee & Herbst (1995) show that complex molecules can still be produced in dense clouds in the presence of rapid neutral-neutral reactions, depending upon which classes of reactions are actually efficient. Given the current uncertainty in the use of rapid neutral-neutral reactions, it is probably best to heed the advice of Bettens et al. and restrict our theoretical comparisons to more standard models for the time being.

Charnley et al. (1992) have modeled the hot core in Orion by following the chemistry in the hot gas after evaporation of the grain mantles. In their model, the composition of the mantles at the time of release into the gas phase is a free parameter. The relevant abundances at $6 \times 10^4$ y after evaporation of the grain mantles are shown in column 9 of Table 7. HCN is produced in the hot-gas phase at an abundance observed in both GL 2591 and the Orion hot core. The good agreement with the $C_2H_2$ abundance is simply due to the assumed initial mantle composition. According to Charnley et al., $C_2H_2$ cannot be formed in the hot-gas phase and there is no obvious grain surface route to its formation. This seems to require that $C_2H_2$ originate in the cold gas phase.

The velocity differences between the molecular absorption components and the molecular cloud indicate that shocks could be present and suggest shock chemistry as a possible explanation of the infrared data. In chemical models of shocked dense clouds by Mitchell (1984), HCN does reach abundances similar to those observed in GL 2591, $\sim 1 - 3 \times 10^{-7}$ of H. However, the abundance of $C_2H_2$ never gets above $10^{-8}$ of H, and then only at $10^4$ y. Hence, similar to the Charnley et al (1992) hot core chemistry, large amounts of HCN could be produced, but explaining the $C_2H_2$ abundance would remain a problem.

In summary, none of the models can match the observed abundances in the hot gas as well as gas-phase models at early time ($\sim 10^5$ y). The infrared limits on the cold gas abundances, however, are lower than the predicted abundances at early time and are consistent with steady-state abundances or less, and the radio data require substantial depletions of CS, HCN, $HCO^+$ and $H_2CO$ with respect to the steady-state models. These observations could be reconciled if gas with the peak early-time abundances of pure gas-phase models was accreted and passively stored on grains, to be later released when the dust and gas were heated.

### 5.4. Comparison to Orion IRc2

In Paper I, we studied absorption lines of $C_2H_2$, HCN, OCS, $NH_3$ and CO towards IRc2 in Orion. We recently discovered an error in the curve-of-growth analysis in that paper, which affects the results for $C_2H_2$, HCN, and OCS. This is discussed more fully in Appendix B where results using the correct curve of growth are presented. The column densities of $C_2H_2$, HCN, and OCS drop by a factor of about 1.5. The conclusions from other, less saturated, bands is unchanged, and the derived temperature, $\sim 140$ K, describing the rotational level populations is also unchanged. The CO column density remains rather uncertain, due to line blending and an uncertain degree of saturation, but indicates abundances of $C_2H_2$ and HCN relative to CO of $\sim 10^{-3}$. Hence, the abundances of $C_2H_2$ and HCN towards GL 2591 and IRc2 are similar, though the total column density towards IRc2 is an order of magnitude larger.

Unlike GL 2591, $CH_4$ and $NH_3$ were clearly detected towards IRc2. The detection of gaseous $CH_4$ was reported in Lacy et al. (1991) with an equivalent width in the R(0) line of $2.9 \times 10^{-2}$ cm$^{-1}$. Assuming a temperature of 140 K for this gas, $N(CH_4) = 2.2 \times 10^{17}$ cm$^{-2}$, giving a $CH_4$ abundance of $1.5 \times 10^{-3}$. This value is probably underestimated due to saturation of the R(0) line. The limit on hot $CH_4$ towards GL 2591 is 10 times greater than the Orion measurement.

In Paper I, the total $NH_3$ column density in Orion was not derived since insufficient lines were detected to determine both temperature and column density, and only column densities in each (J,K) level were given. If we assume that a temperature of 140 K is appropriate for the three measured $NH_3$ lines, then $N(NH_3) = (1.7 \pm 0.8) \times 10^{16}$ cm$^{-2}$. The corresponding $NH_3$ abundance is $1.1 \times 10^{-4}$. The tentative detection of cold $NH_3$ in GL 2591 gives a similar abundance, but the limit on $NH_3$/CO in the hot gas is only $10^{-2}$.

The IRc2 abundances derived from the infrared lines are summarized in Table 7. A comparison of the IRc2 abundances to chemical models yields conclusions similar to those reached for GL 2591: the abundances of $C_2H_2$, HCN, and $NH_3$ are best matched by gas-phase chemistry at early times. The abundance of $CH_4$ compares better with steady-state abundances, but a substantial correction for saturation could produce agreement with early-time abundances. Hence, freeze out of gas with high peak abundances, with little or no grain processing, and later release into the gas is a consistent model for the infrared absorption measurements of IRc2 and GL 2591. In Orion, additional evidence for this scenario comes from the large abundances of deuterated molecules (e.g., Walmsley et al. 1987; Mauersberger et al. 1988; Mangum et al. 1991), though the detection of doubly deuterated formaldehyde (Turner 1990) strongly indicates an active grain-surface chemistry. Radio measurements of $NH_3$ in the hot core give a larger $NH_3$ abundance than the infrared observations, $\sim 10^{-3}$ of CO (see van Dishoeck et al. 1993) vs. $\sim 10^{-4}$, and would also require active grain chemsitry.

## 6. CONCLUSIONS

We have obtained infrared absorption spectra of HCN and $C_2H_2$ towards GL 2591. Our analysis of the data indicate that most of the absorption arises in gas at 200 K and 1010 K, corresponding to the absorption components observed in CO by Mitchell et al. (1989). The abundances of HCN and $C_2H_2$ are about $1 - 3 \times 10^{-3}$ with respect to CO. Cold HCN associated with the 38 K CO absorption component is not observed, but cold $C_2H_2$ is tentatively detected from our synthesis of the $C_2H_2$ Q branch, giving an abundance of $1 \times 10^{-4}$, about 10 times lower than in the hot gas. Radio lines of HCN, CS, $H_2CO$ and $HCO^+$ were also observed. Analysis of the radio data gives an abundance of about $7 \times 10^{-6}$ for HCN relative to CO, more than two orders of magnitude below the infrared absorption measurements. The radio data also indicate that the cloud surrounding GL2591 has a density gradient with a power-law exponent of about 1.5.

We have determined the densities in the hot gas to be $\sim 3 \times 10^7$ cm$^{-3}$, based on a non-LTE analysis of the HCN rovibrational lines. A similar density is obtained by extrapolating our spherical cloud models for the radio emission inward to a radius of $1''$. This radius is consistent with our estimate for the maximum extent of 2-3$''$ for the hot infrared absorbing gas obtained by comparing the radio and infrared data.

The radio emission lines and the infrared absorption lines are clearly not produced in the same gas. Besides the differences in temperature and HCN abundance, the absorption lines, with $V_{LSR} = -8$ to $-28$ km s$^{-1}$, are all blueshifted with respect to the $V_{LSR} = -5.8$ km s$^{-1}$ peak of the radio emission lines. In fact, the absence of a strong CO absorption component at $-5.8$ km s$^{-1}$ (as well as the relatively low extinction toward the source) indicates that GL 2591 is being viewed through a "hole" in the ambient molecular cloud, perhaps an opening caused by the outflow, though modeling of the radio lines indicates that the hole cannot be spherically symmetric about the source. Near-infrared images indicate a shell-like cavity, probably created by the outflow.

A comparison with chemical models for dense molecular clouds shows that the abundances of HCN and $C_2H_2$ in the hot gas are most similar to models using pure gas-phase chemistry at $t \sim 10^5$ y. On the other hand, the $C_2H_2$ abundance in the cold gas is consistent with steady-state gas-phase models; the limit on $CH_4$ is also far lower than the models at early times, and our marginal detection of $NH_3$ is lower than the steady-state predictions. The abundances of HCN, CS, $H_2CO$ and $HCO^+$ derived from modeling the radio data are lower than steady-state models by factors of 2-30, indicating some depletion of molecules in the ambient cloud. Models which include grain-surface chemistry do not explain the infrared data: depending on the model, abundances of HCN, $NH_3$, or $CH_4$ are too high. A scenario in which grains play a passive role can explain the combined data. Early in the history of the molecular cloud, molecules with abundances which peak at $\sim 10^5$ y in the gas-phase chemistry are frozen onto the grains. Following star formation and heating of the gas and dust, these molecules are released into the hot gas. The radio and infrared observations of the cold gas reveal gas abundances which are characteristic of later times ($> 10^6$ y) with some gas-phase depletion. The hot gas observed in absorption is likely to be material accelerated by the outflow. This gas has been heated by either radiation or shocks, sublimating the grain mantles and releasing gas which has abundances characteristic of early-time gas-phase chemistry.

JC acknowledges support from NSF grant AST86-15906 and from a Columbus Fellowship at The Ohio State University. NJE acknowledges support from NSF grant AST93-17567 and helpful discussions with M. Choi about the Monte Carlo code. Observations with Irshell were supported by NSF grant AST90-20292.

## A. SPECTROSCOPIC DATA

In Paper I we presented the basic equations and spectroscopic data used to analyze the infrared molecular spectra. Here we give the basic data for molecules which were not studied in Paper I.

Both CS and SiO are diatomic molecules with no residual electronic angular momentum, and the Hönl-London factors are given by equation (A10) in Paper I. The line frequencies and molecular parameters for CS are taken from Burkholder et al. (1987b). Botschwina & Sebald (1985) calculate $|\mu_v| = 0.158$ debye for CS $v = 1 - 0$, from which we calculate a band strength $S = S^i_\nu = 1.32 \times 10^{-17}$ cm$^{-1}$/ cm$^{-2}$. Frequencies and constants for SiO are from Lovas, Maki & Olson (1981). We take $|\mu_v| = 0.094$ debye from the calculations for SiO of Tipping & Chackerian (1981), which gives $S = S^i_\nu = 0.453 \times 10^{-17}$ cm$^{-1}$/ cm$^{-2}$.

The line frequencies and molecular parameters for SO are from Burkholder et al. (1987a). We observed rovibrational transitions of the ground electronic state, $X^3\Sigma^-$. Tatum (1966) has tabulated the Hönl-London factors for such transitions.

The total column densities in $NH_3$ towards GL 2591 and IRc2 were calculated from the line equivalent widths for the assumed temperature using

$$N = \frac{\nu_i}{\nu_l} \frac{w_\nu Q_r(T)}{s_K L(J) S} \exp(E_l/kT), \tag{A1}$$

where $\nu_i$ is the band center frequency, $\nu_l$ the line frequency, $Q_r(T)$ the rotational partition function, and $s_K$ is the statistical weight due to nuclear spin ($s_K = 4$ for K = multiple of 3 or K = 0, and $s_K = 2$ for K not a multiple of 3) (Townes & Schawlow 1955). For this convention for the $s_K$ and for $kT >> hB$, the rotational partition function can be written as

$$Q_r(T) \simeq 2.67(\frac{\pi k^3}{B^2 C h^3})^{1/2} T^{3/2}, \tag{A2}$$

where B and C are the rotational constants in Hz. Note that other conventions are in common use; Gordy & Cook (1970) give $s_K$ and $Q_r$ which are both 1/8 of the values given here. The Hönl-London factors are given by equation (A25) in Paper I, and Table 9 in Paper I gives the other molecular data.

## B. Revised Curve of Growth Analysis for Orion IRc2

In a previous paper (Evans et al. 1991) we used infrared absorption line observations of $C_2H_2$, HCN, OCS, and $NH_3$ to study the gas surrounding the sources IRc2 and IRc7 in Orion. The relative strengths of the ortho and para line of $C_2H_2$ implied line-center optical depths greater than one, and we used an analytic approximation to the curve of growth of a Gaussian line to fit the measured equivalent widths.

Since the publication of our Orion paper, it was pointed out to us by S. Willner that the approximation we used is not valid for optical depths as large as those implied by the analysis. The assumed curve of growth more nearly approximated that of an exponential ($exp[-(\nu - \nu_0)/\Delta\nu]$) lineshape than that of a Gaussian at large optical depths, and it continued to rise at optical depths where a Gaussian line curve of growth is essentially level. Consequently, we repeated the analysis of the Orion data using a numerically integrated curve of growth. The main change that results from the corrected curve-of-growth analysis is a decrease in the $C_2H_2$ column density derived from the $\nu_5$ band of $C_2H_2$ and in the HCN and OCS column densities by factors of order two. The new column densities are given in Table 9. This change is not large compared to the uncertainties, especially in the comparison of the observed column densities with the estimated CO and $H_2$ column densities.

A possibly more interesting effect of the corrected curve of growth is an increased disagreement between the $C_2H_2$ column densities derived from the $\nu_5$ (13.7$\mu$m) and the $\nu_4 + \nu_5$ (7.5 $\mu$m) bands. With the erroneous curve of growth, the $\nu_4 + \nu_5$ band required $\sim 3$ times more $C_2H_2$ than did the $\nu_5$ band, whereas with the numerical curve of growth the $\nu_5$ band requires $\sim 5 - 6$ times more $C_2H_2$. Two explanations for the now quite large discrepancy between the two bands seem most reasonable. First, as suggested by Evans et al., the two bands may sample different gas. The continuum against which we see the $C_2H_2$ absorption is formed by dust emission. The continuum will form farther from IRc2 at 13.5 $\mu$m than at 7.5 $\mu$m for two reasons. The optical depth of interstellar dust is about twice as large at 13.5 $\mu$m as it is at 7.5 $\mu$m (Draine 1989). In addition, and possibly more significantly, the shorter wavelength continuum is biased toward hotter grains, which will be closer to IRc2. For this effect to explain the discrepancy, most of the $\nu_4 + \nu_5$ $C_2H_2$ absorption would have to arise in gas inside the photosphere at 13.5 $\mu$m. An argument against this explanation is the fact that unsaturated $^{13}C^{12}CH_2$ lines were observed and require a column density of $^{13}C^{12}CH_2$ only 16 times smaller than that of $^{12}C_2H_2$, or an atomic $^{12}C/^{13}C$ abundance ratio of 32. The previous curve of growth gave a more acceptable $^{12}C/^{13}C$ ratio of 50. Since the $^{13}C^{12}CH_2$ lines are at essentially the same wavelength as the $C_2H_2$ $\nu_5$ lines, both bands must sample the same gas.

A second possibility is that the actual lineshapes are not Gaussian. To lessen the disagreement between the two bands, a larger saturation correction would have to be made for the $\nu_5$ band. Since it is the slope of the curve of growth that is determined by the ortho-to-para ratio, a softer curve of growth, resulting from a lineshape with stronger wings, would require a larger degree of saturation. The erroneous analytic curve of growth differed from a Gaussian curve of growth in this way, but not by enough to eliminate the discrepancy between the two bands. The differences that result from the use of different intrinsic line shapes underscores the uncertainty in the curve-of-growth analysis.

Although both explanations for the disagreement between the different $^{12}C_2H_2$, bands have problems, we find the first more plausible, that the absorbing gas is mixed with the emitting dust, and that observations at different wavelengths probe different columns of material. We note that although there is no evidence of this problem in the case of GL 2591 (since we were unable to detect either the $\nu^4 + \nu^5$ or the $^{13}C^{12}CH_2$ bands), similar effects could easily occur.

Table 1. Observations of $C_2H_2$

| Line | $w_\nu$ ($10^{-3}$ cm$^{-1}$) | $V_{LSR}$ (km s$^{-1}$) |
|---|---|---|
| R(1) | $1.4 \pm 0.5$ | $-15.4$ |
| R(5) | $6.9 \pm 0.5$ | $-23.6$ |
| R(10) | $2.9 \pm 1.0$ | $-14.1$ |
| R(14) | $1.7 \pm 0.5$ | $-7.5$ |
| R(15) | $7.4 \pm 1.0$ | $-14.2$ |
| R(19) | $4.3 \pm 1.4$ | $-11.0$ |
| R(21) | $5.0 \pm 1.0$ | $-10.0$ |
| R(19)[a] | $4.6 \pm 1.0$ | ··· |
| R(21)[a] | $3.5 \pm 0.6$ | ··· |
| R(0) $\nu_4 + \nu_5$ | $< 1.3$[b] | ··· |

[a]Measurements made at higher dispersion.

[b]Upper limit (2 $\sigma$).

Table 2. Observations of HCN

| Line | $W_\nu$ ($10^{-3}$ cm$^{-1}$) | $V_{LSR}$ (km s$^{-1}$) |
|---|---|---|
| R(6) | $7.1 \pm 1.1$ | −3.4 |
| R(10) | $6.6 \pm 1.0$ | −15.7 |
| R(14) | $3.3 \pm 0.6$ | −11.2 |
| R(17) | $3.0 \pm 0.5$ | −3.2 |
| R(21) | $1.9 \pm 1.0$ | −11.3 |
| R(22) | $1.5 \pm 1.0$ | −7.0 |
| R(21)[a] | $< 0.5$[b] | ··· |

[a]Measurements made at higher dispersion.

[b]Upper limit (2 $\sigma$).

Table 3. Limits on Other Molecules

| Molecule | Transition | $w_\nu$ ($10^{-3}$ cm$^{-1}$) |
|---|---|---|
| $CH_4$[a] | R(0) | $< 8.0$[b] |
| $NH_3$ | aR(0,0) | $\leq 1.5$ |
| | sP(3,K) | $< 0.8$ |
| | sQ | $< 0.8$ |
| CS | R(7),R(11) | $< 1.5$ |
| SiO | R(0),R(1),P(15) | $< 1.4$ |
| SO | R(0),R(1),$R_3$(2) | $< 1.2$ |

[a]Lacy et al. (1991).

[b]Upper limits (2 $\sigma$).

Table 4. Observational Parameters for Radio Data

| Species | Line | $\nu$(GHz) | Telescope | $\theta_b$ | $\eta_{mb}$ | RMS | Cal.[a] | Map? | No. Points |
|---|---|---|---|---|---|---|---|---|---|
| CS | $2-1$ | 97.980968 | IRAM | 24 | 0.60 | 0.25 | 20 | no | ··· |
| CS | $3-2$ | 146.969049 | IRAM | 17 | 0.55 | 0.52 | 50 | no | ··· |
| CS | $5-4$ | 244.935606 | IRAM | 11 | 0.45 | 0.54 | 20 | no | ··· |
| CS | $5-4$ | 244.935610 | CSO | 30 | 0.72 | 0.17 | 20 | yes | 17 |
| CS | $7-6$ | 342.882900 | CSO | 20 | 0.55 | 0.05 | 20 | yes | 11 |
| CS | $10-9$ | 489.751040 | CSO | 14 | 0.39 | 0.23 | 30 | no | ··· |
| $C^{34}S$ | $2-1$ | 96.412982 | IRAM | 24 | 0.60 | 0.12 | 20 | no | ··· |
| $C^{34}S$ | $3-2$ | 144.617147 | IRAM | 17 | 0.55 | 0.13 | 20 | no | ··· |
| $C^{34}S$ | $5-4$ | 241.016176 | CSO | 30 | 0.55 | 0.07 | 20 | no | ··· |
| HCN | $3-2$ | 265.886180 | CSO | 27 | 0.72 | 0.06 | 20 | yes | 18 |
| HCN | $4-3$ | 354.505470 | CSO | 20 | 0.55 | 0.07 | 20 | yes | 26 |
| $H^{13}CN$ | $3-2$ | 259.011828 | CSO | 27 | 0.72 | 0.05 | 20 | no | ··· |
| $HCO^+$ | $3-2$ | 267.557625 | CSO | 26 | 0.72 | 0.12 | 20 | yes | 21 |
| $H^{13}CO^+$ | $4-3$ | 346.998540 | CSO | 20 | 0.65 | 0.10 | 20 | yes | 13 |
| $H_2CO$ | $3_{12}-2_{11}$ | 225.697780 | CSO | 31 | 0.72 | 0.20 | 20 | no | ··· |
| $H_2CO$ | $5_{15}-4_{14}$ | 351.769000 | CSO | 20 | 0.55 | 0.07 | 20 | yes | 26 |

[a]Calibration uncertainty in percent.

Table 5. Derived Column Densities

| Molecule | $N_{38K}$ ($10^{16}$ cm$^{-2}$) | $N_{200K}$ ($10^{16}$ cm$^{-2}$) | $N_{1010K}$ ($10^{16}$ cm$^{-2}$) |
|---|---|---|---|
| $C_2H_2$ | $\leq 0.08$[a] | $0.42 \pm 0.07$ | $1.05 \pm 0.16$ |
| HCN | $< 1.7$[b] | $2.0 \pm 0.4$ | $1.6 \pm 0.5$ |
| CO[c] | $720 \pm 120$ | $660 \pm 180$ | $560 \pm 110$ |
| $CH_4$ | $< 0.8$ | $< 10$ | $< 130$ |
| $NH_3$ | $\leq 0.05$ | $< 0.1$ | $< 0.7$ |
| CS | $< 0.26$ | $< 0.34$ | $< 0.9$ |
| SiO | $< 0.6$ | $< 1.6$ | $< 2.5$ |

[a]From the analysis of the Q branch.

[b]Upper limits (2 $\sigma$).

[c]The CO column densities are from Mitchell et al. (1989).

Table 6. Abundances Relative to CO from Infrared – Cold Gas

| Molecule | Observed | Gas Models[a] | | Gas–Grain Models | | | |
|---|---|---|---|---|---|---|---|
| | | $10^5$ y | Steady State | $10^5$ y[b] | $10^6$ y[b] | B90[c] | BCM[d] |
| $C_2H_2$ | $\leq 1(-4)$ | 1.0(−2) | 1.0(−4) | 3.9(−3) | 2.8(−3) | ··· | 1.9(−2) |
| HCN | $< 2(-3)$ | 1.9(−3) | 1.1(−5) | 1.2(−3) | 2.0(−4) | ··· | 1.7(−2) |
| $CH_4$ | $< 1(-3)$ | 5.7(−2) | 1.5(−3) | 2.6(−2) | 1.1(−2) | ··· | 7.0(−3) |
| $NH_3$ | $\leq 7(-5)$ | 3.2(−4) | 3.5(−4) | 2.2(−4) | 1.2(−3) | ··· | 1.6(−4) |
| CS | $< 4(-4)$ | 1.2(−4) | 8.7(−5) | 1.0(−4) | 4.0(−5) | ··· | ··· |
| SiO | $< 8(-4)$ | 9.1(−6) | 4.7(−6) | 7.8(−6) | 4.5(−6) | ··· | ··· |
| $H_2O$ ice | 0.24 | 0.03[e] | 0.009[e] | 0.75 | 12.1 | 1.5 | 7.0 |
| CO ice | $< 0.004$[f] | ··· | ··· | 0.04 | 0.53 | ··· | ··· |
| $CO/H_2$ | 1(−4)[g] | 8(−5) | 1.5(−4) | 6.7(−5) | 1.4(−5) | 5.8(−5) | 1.3(−4) |

Note. — $a(-b) = a \times 10^{-b}$

[a]Gas-phase models of Herbst & Leung 1989, standard rates, low metals.

[b]Gas-phase abundances from Hasegawa & Herbst 1993, Model N(2100K, CR).

[c]Brown 1990, Case 3.

[d]Brown et al. 1988, Case A; gas at $t \sim 4 \times 10^5$ y, water ice at $t \sim 10^6$ y.

[e]Abundance assuming all water frozen on grains.

[f]From Mitchell et al. 1990.

[g]Using $N(H_2)$ estimated from the silicate feature.

Table 7. Abundances Relative to CO from Infrared – Hot Gas

| Molecule | Observed | Gas Models[a] | | Gas–Grain Models | | | | CTM[e] | IRc2 |
|---|---|---|---|---|---|---|---|---|---|
| | | $10^5$ y | Steady State | $10^5$ y[b] | $10^6$ y[b] | B90[c] | BCM[d] | | |
| $C_2H_2$ | 1(−3) | 5.4(−3) | 1.0(−4) | 4.8(−3) | 5.8(−3) | 9.4(−3) | 1.7(−2) | 3.2(-3) | 6(−4) |
| HCN | 3(−3) | 1.0(−3) | 1.1(−5) | 6.2(−2) | 0.28 | 1.0(−2) | 1.3(−3) | 2.7(-3) | 1(−3) |
| $CH_4$ | $<$ 1.5(−2) | 3.1(−2) | 1.5(−3) | 3.2(−2) | 0.13 | 4.9(−3) | 1.64 | 5.0(-3) | $\geq$ 1.5(−3) |
| $NH_3$ | $<$ 1(−2) | 1.7(−4) | 3.5(−4) | 3.1(−4) | 1.8(−3) | 0.23 | 1.64 | 1.5(-2) | 1.1(−4) |
| CS | $<$ 5(−4) | 6.5(−5) | 8.7(−5) | 9.8(−5) | 3.3(−5) | ··· | ··· | ··· | ··· |
| SiO | $<$ 2(−3) | 4.9(−6) | 4.7(−6) | 9.6(−6) | 2.1(−5) | ··· | ··· | ··· | ··· |
| $CO/H_2$ | ··· | 8.1(−5) | 1.5(−4) | 6.9(−5) | 2.1(−5) | 3.2(−4) | 1.3(−4) | ··· | ··· |

Note. — $a(-b) = a \times 10^{-b}$

[a]Abundances from standard model of Herbst & Leung 1989 with respect to steady-state CO abundance.

[b]Gas + surface abundances from Hasegawa & Herbst 1993, Model N(2100K, CR).

[c]Brown 1990, Case 3.

[d]Brown et al. 1988, Case A; final abundances after grain mantle evaporation.

[e]Hot core model of Charnley et al. 1992 at $t = 6.3 \times 10^4$ y after grain evaporation.

Table 8. Abundances ($10^{-10}$) Relative to $H_2$ from Radio and Model Predictions

| Species | $X(1.5)$[a] | HLSS[b] | LG[c] | HH6[d] | HHCR6[e] | HHCR7[f] |
|---|---|---|---|---|---|---|
| CS | 4 | 130 | 50 | 5.1 | 5.6 | $1.9 \times 10^{-5}$ |
| HCN | 7 | 16 | 300 | 11.0 | 28.0 | 1.8 |
| $HCO^+$ | 4 | 95 | 60 | 64.0 | 66.0 | 0.008 |
| $H_2CO$ | 2 | 33 | 80 | 77.0 | 90.0 | 0.27 |

[a]Best-fitting abundance for $\alpha = 1.5$.

[b]Abundance from Herbst & Leung 1989 for standard rates, low metals, and steady state.

[c]Abundance from Langer & Graedel 1989 for $t = 10^8$ yr, $n = 5 \times 10^3$ cm$^{-3}$, and $T = 10$ K.

[d]Gas-phase abundance from Hasegawa & Herbst 1993 for $t = 10^6$ yr, normal initial abundances, without cosmic ray desorption.

[e]Gas-phase abundance from Hasegawa & Herbst 1993 for $t = 10^6$ yr, normal initial abundances, with cosmic ray desorption.

[f]Gas-phase abundance from Hasegawa & Herbst 1993 for $t = 10^7$ yr, normal initial abundances, with cosmic ray desorption.

Table 9. Results for Orion IRc2 with New Curve of Growth

| Species | Band | Branch | T (K) | N ($10^{17}$ cm$^{-2}$) | $\sigma_v$ [a] (km s$^{-1}$) |
|---|---|---|---|---|---|
| $C_2H_2$ | $\nu_5$ | R | 152 | 0.95 | 2.0 |
| $^{13}C_2H_2$ | $\nu_5$ | R | 67 | 1.5 | (thin) |
| $C_2H_2$ | $\nu_4 + \nu_5$ | P, R | 159 | 5.0[b] | 2.6 |
| HCN | $\nu_2$ | R | 132 | 1.8 | (2.0) |
| OCS | $\nu_1$ | P, R | 134 | 0.21 | (2.0) |

[a]Linewidths in parentheses are assumed.

[b]The column density is that of $^{13}C_2H_2$ multiplied by 25, which assumes an $^{12}C/^{13}C$ ratio of 50.

---

Fig. 1.— Reduced spectra of the observed 13$\mu$m lines of $C_2H_2$ and HCN, and the aR(0,0) line of $NH_3$. The data are not plotted when the atmospheric transmission is less than 80 % of the local peak transmission. Some additional residual telleric lines are marked.

Fig. 2.— Q branch of $C_2H_2$ in GL 2591. a) The atmospheric transmission (as determined from the difference between spectra of the atmospheric emission and a blackbody) with the position of some telluric water lines marked. b) The ratioed spectrum of the Q branch with the best single-temperature fit from Fig. 4. The histogram is the data and the smooth line is the model. The wavelengths for individual transitions are marked for $V_{LSR} = -14$ km s$^{-1}$, with only the odd-J lines labeled. The HCN R(5) line was not included. c) Q branch with the best two-temperature fit from Fig. 5 and the prediction for the HCN R(5) line. The wavelengths of the two velocity components of the HCN R(5) line are marked. d) Q branch with three-temperature model from Fig. 5 with $N_{38} = 6 \times 10^{14}$ cm$^{-2}$(solid line), including the HCN R(5) line; the two dashed lines show the effect of increasing $N_{38}$ to $10 \times 10^{14}$ cm$^{-2}$and $30 \times 10^{14}$ cm$^{-2}$.

Fig. 3.— Maps of $T_A^*$ made at the CSO for (top row) CS $J = 5-4$ and $J = 7-6$, (second row) HCN $J = 3-2$ and $J = 4-3$, and (third row) $HCO^+$ $J = 3-2$ and $H^{13}CO^+$ $J = 4-3$. The resolution of the maps in the left column is about 26 to 30$''$, while that of the maps in the right column is about 20$''$. The contours begin at $2\sigma$ and are spaced by $2\sigma$; contour intervals are noted at the bottom of each panel.

Fig. 4.— The natural logarithm of the column density of $C_2H_2$ per degenerate sublevel versus the energy of the lower level. The straight line is the best linear fit to the data and gives of temperature of 410 $\pm$ 40 K and a column density in $C_2H_2$ of $(1.2 \pm 0.2) \times 10^{16}$ cm$^{-2}$. $\chi_r^2 = 1.07$ for the fit.

Fig. 5.— Same as Fig. 4, but the solid line now shows the best two-temperature fit, with $N_{200} = (4.2 \pm 0.7) \times 10^{15}$ cm$^{-2}$and $N_{1010} = (1.05 \pm 0.16) \times 10^{16}$ cm$^{-2}$. The dashed line is the 2 $\sigma$ limit on cold $C_2H_2$, with $N_{38} = 6.0 \times 10^{14}$ cm$^{-2}$, $N_{200} = 3.0 \times 10^{15}$ cm$^{-2}$and $N_{1010} = 1.1 \times 10^{16}$ cm$^{-2}$.

Fig. 6.— Same as Fig. 4, but for HCN. The straight line is a linear fit to the four lowest rotational lines and gives a temperature of 303 $\pm$ 38 K and a HCN column density of $(3.1 \pm 0.6) \times 10^{16}$ cm$^{-2}$. $\chi_r^2 = 1.18$. The inverted triangle is the 2 $\sigma$ limit on the R(21) line; the R(22) line is only a 1.5 $\sigma$ number and was treated as an upper limit.

Fig. 7.— Same as Fig. 5, but for HCN. The solid line shows the best two-temperature fit to the four lowest rotational lines, with $N_{200} = (2.0 \pm 0.4) \times 10^{16}$ cm$^{-2}$and $N_{1010} = (1.6 \pm 0.5) \times 10^{16}$ cm$^{-2}$. $\chi_r^2 = 0.63$. The dashed line is the 2 $\sigma$ limit on cold HCN, with $N_{38} = 1.7 \times 10^{16}$ cm$^{-2}$, $N_{200} = 1.3 \times 10^{16}$ cm$^{-2}$and $N_{1010} = 2.1 \times 10^{16}$ cm$^{-2}$. The inverted triangle is the 2 $\sigma$ limit on the R(21) line; the R(22) line is only a 1.5 $\sigma$ number and was treated as an upper limit.

Fig. 8.— The HCN populations per sublevel are plotted versus energy as in previous figures. The three lines are the combination of populations in two temperature components, one at 200 K and one at 1010 K. The density was assumed to be the same in both temperature components; three different densities are shown to give an idea of the sensitivity to density. The column densities of each component were adjusted independently, but the best solution was with equal column densities The inverted triangle is the 2 $\sigma$ limit on the R(21) line; the R(22) line is only a 1.5 $\sigma$ number and was treated as an upper limit.

Fig. 9.— The radio data for CS and $C^{34}S$ are shown along with the best-fitting model (dashed line) for $\alpha = 1.5$. The first row is the IRAM data, expressed at $T_{mb}$. The second row is the CSO data, expressed as $T_A^*$; the models have incorporated the telescope efficiency, so are directly comparable to the data. The third row is $C^{34}S$ data; the first two panels are IRAM data, expressed in $T_{mb}$, while the last panel is CSO data, expressed in $T_A^*$.

Fig. 10.— The radio data for HCN, $H_2CO$, and $HCO^+$, as well as off-center positions for HCN, CS, and $HCO^+$ are shown, along with the best-fitting model (dashed line), for $\alpha = 1.5$. All are CSO data, expressed as $T_A^*$; the models have incorporated the telescope efficiency, so are directly comparable to the data. The first row is the HCN and $H^{13}CN$ data toward the center position. The second row shows the two HCN transitions at positions offset by $18''$ and the $H_2CO$ $J = 5-4$ line toward the center. Data at the offset positions were obtained by averaging four spectra displaced in the four cardinal directions for the $J = 3-2$ line; for the $J = 4-3$ line, data were averaged in such a way as to remove the effect of the pointing shift seen in the $J = 4-3$ map. The third row shows CS data at offset positions and the $H_2CO$ $J = 3-2$ line at the center position. The fourth line shows $HCO^+$ and $H^{13}CO^+$ (multiplied by 5) data at the center and $HCO^+$ data (multiplied by 4) at a position offset by $36''$.

Fig. 11.— The radio data for CS and $C^{34}S$ are shown along with the best-fitting model (dashed line) for $\alpha = 1.0$. The first row is the IRAM data, expressed at $T_{mb}$. The second row is the CSO data, expressed as $T_A^*$; the models have incorporated the telescope efficiency, so are directly comparable to the data. The third row is $C^{34}S$ data; the first two panels are IRAM data, expressed in $T_{mb}$, while the last panel is CSO data, expressed in $T_A^*$.

Fig. 12.— The radio data for HCN, $H_2CO$, and $HCO^+$, as well as off-center positions for HCN, CS, and $HCO^+$ are shown, along with the best-fitting model (dashed line), for $\alpha = 1.0$. All are CSO data, expressed as $T_A^*$; the models have incorporated the telescope efficiency, so are directly comparable to the data. The first row is the HCN and $H^{13}CN$ data toward the center position. The second row shows the two HCN transitions at positions offset by $18''$ and the $H_2CO$ $J = 5-4$ line toward the center. Data at the offset positions were obtained by averaging four spectra displaced in the four cardinal directions for the $J = 3-2$ line; for the $J = 4-3$ line, data were averaged in such a way as to remove the effect of the pointing shift seen in the $J = 4-3$ map. The third row shows CS data at offset positions and the $H_2CO$ $J = 3-2$ line at the center position. The fourth line shows $HCO^+$ and $H^{13}CO^+$ (multiplied by 5) data at the center and $HCO^+$ data (multiplied by 4) at a position offset by $36''$.

Fig. 13.— As for figure 10, but showing the best model with $\alpha = 2$.

Fig. 14.— As for figure 11, but showing the best model with $\alpha = 2$.

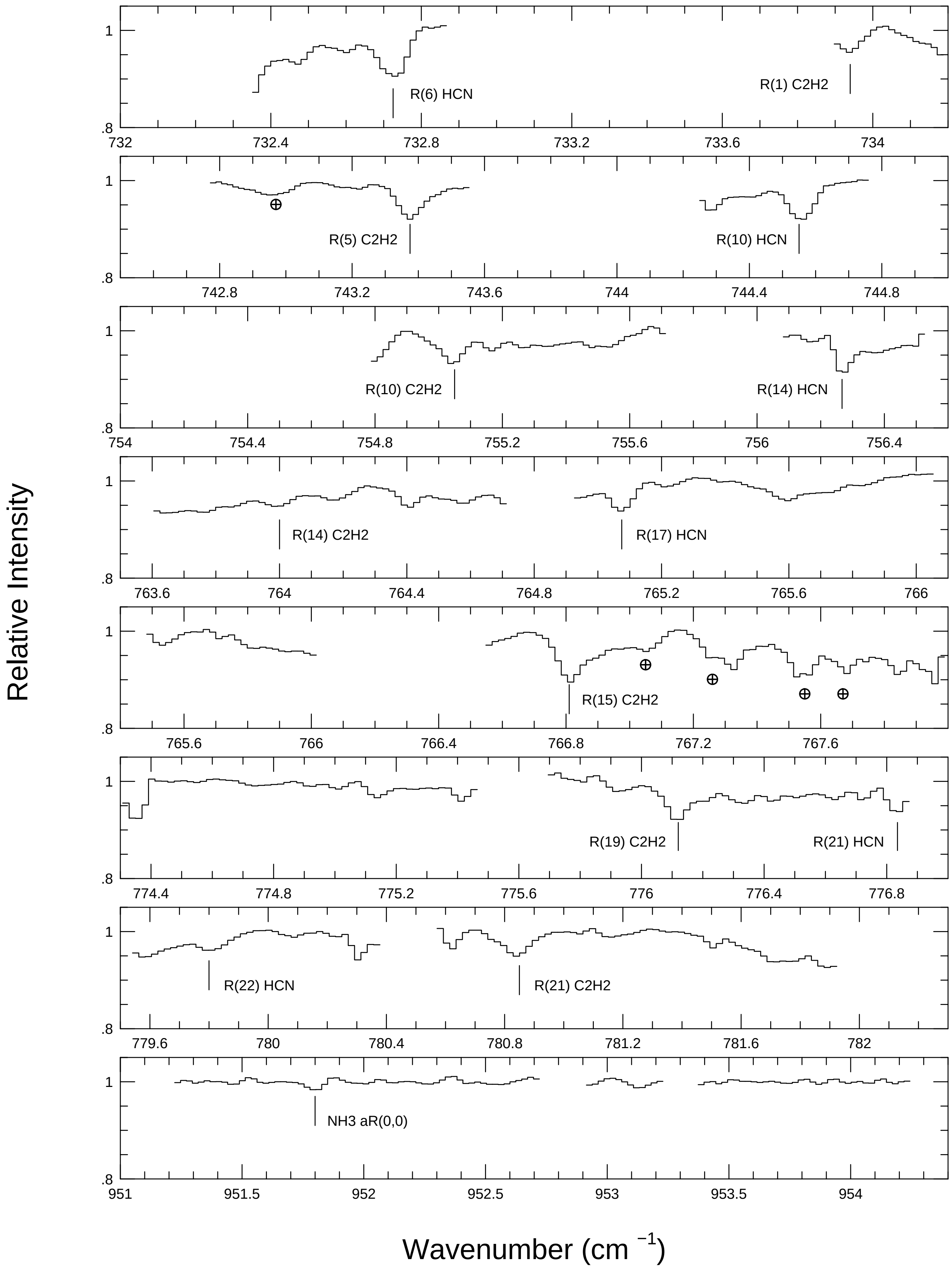

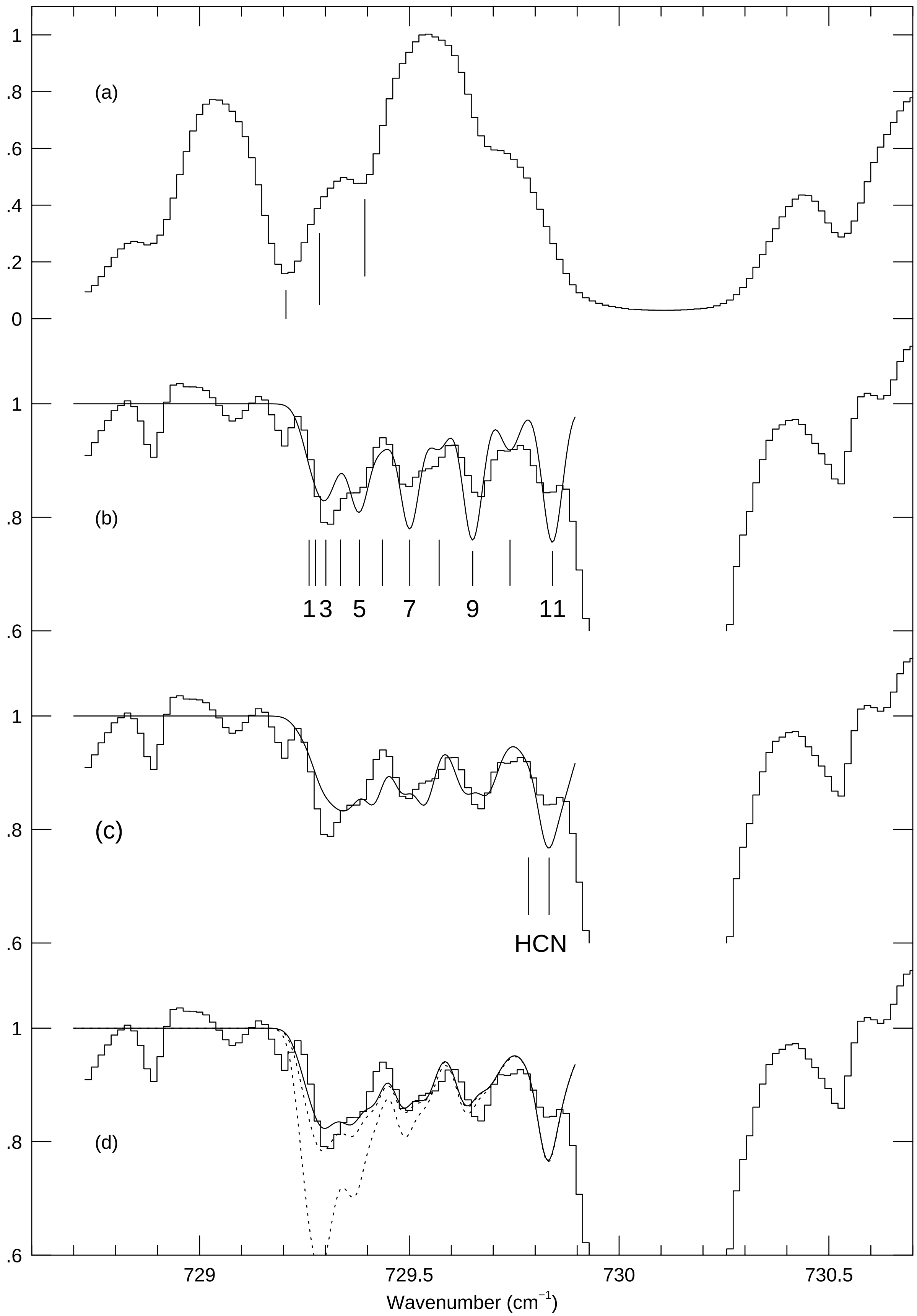

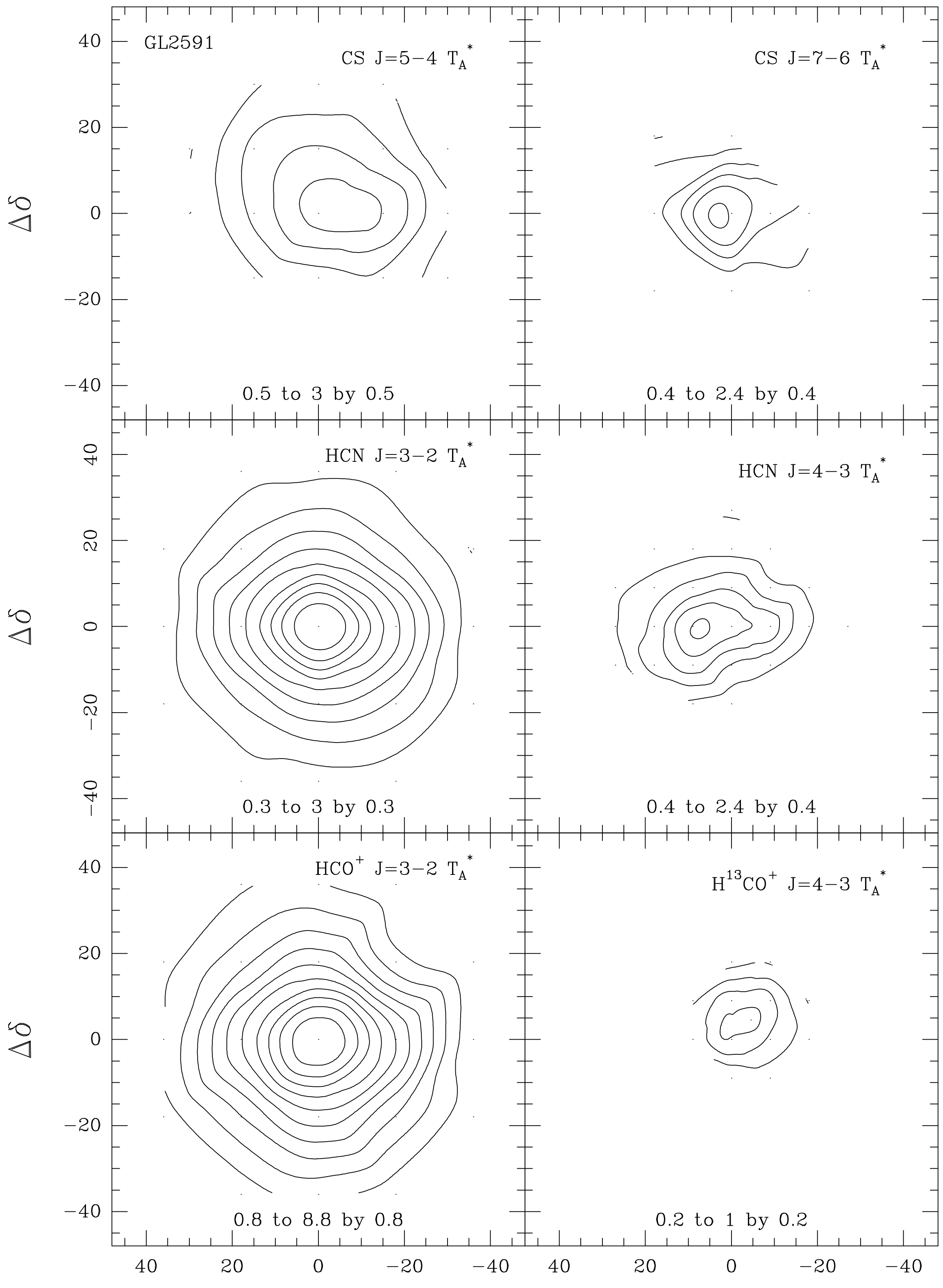
GL2591
CS J=5−4 $T_A^*$
0.5 to 3 by 0.5
CS J=7−6 $T_A^*$
0.4 to 2.4 by 0.4
HCN J=3−2 $T_A^*$
0.3 to 3 by 0.3
HCN J=4−3 $T_A^*$
0.4 to 2.4 by 0.4
$HCO^+$ J=3−2 $T_A^*$
0.8 to 8.8 by 0.8
$H^{13}CO^+$ J=4−3 $T_A^*$
0.2 to 1 by 0.2
$\Delta\delta$
$\Delta\alpha$

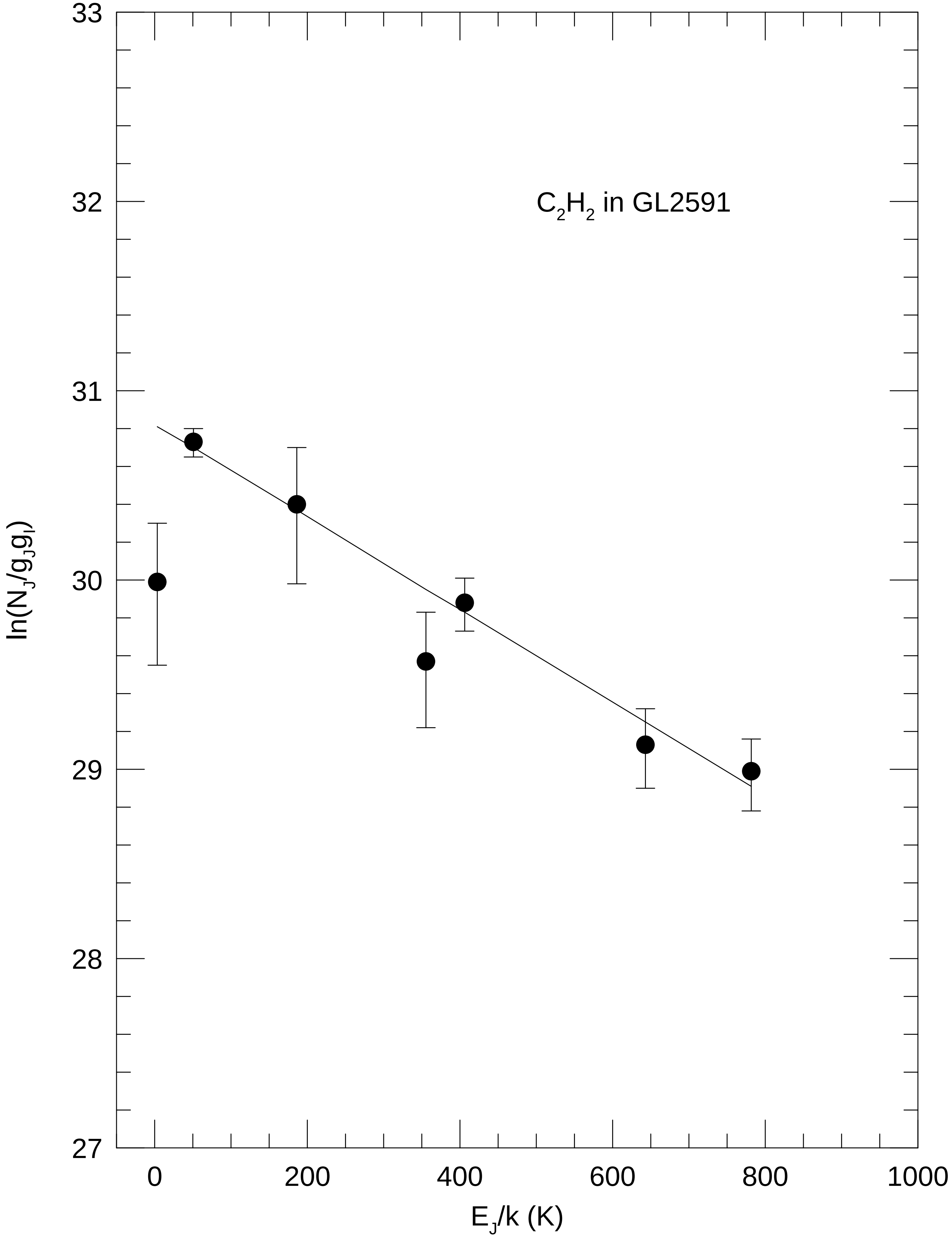

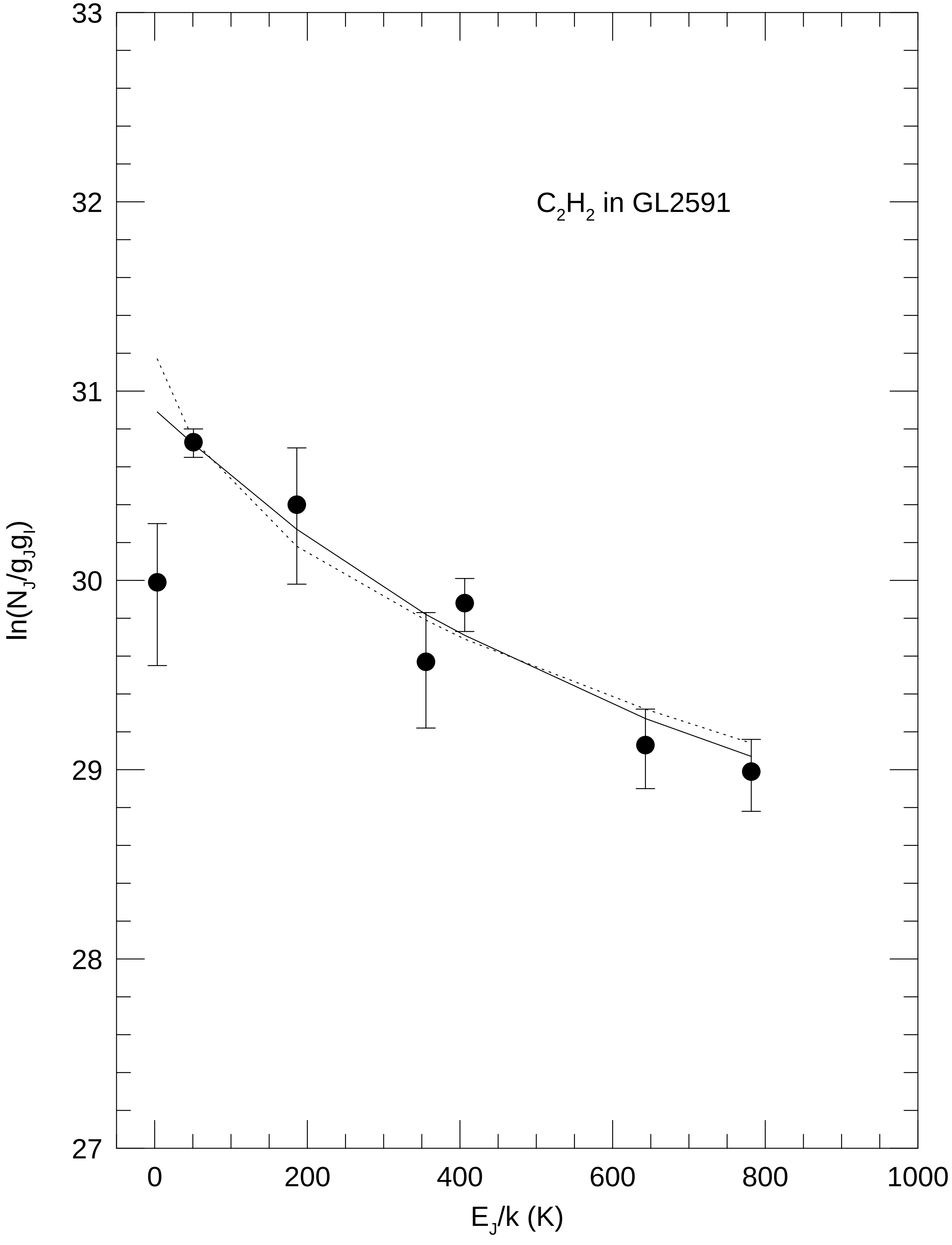

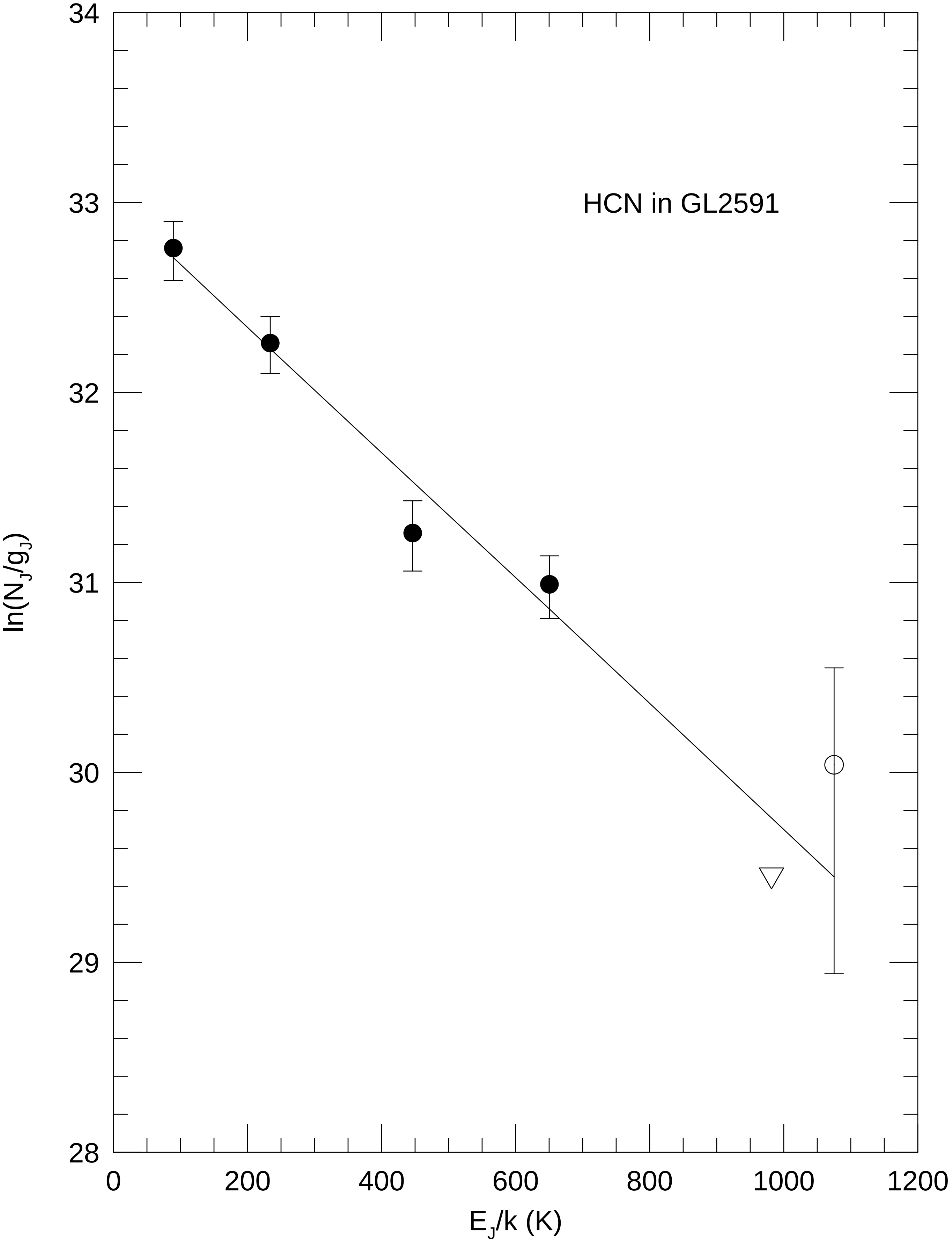

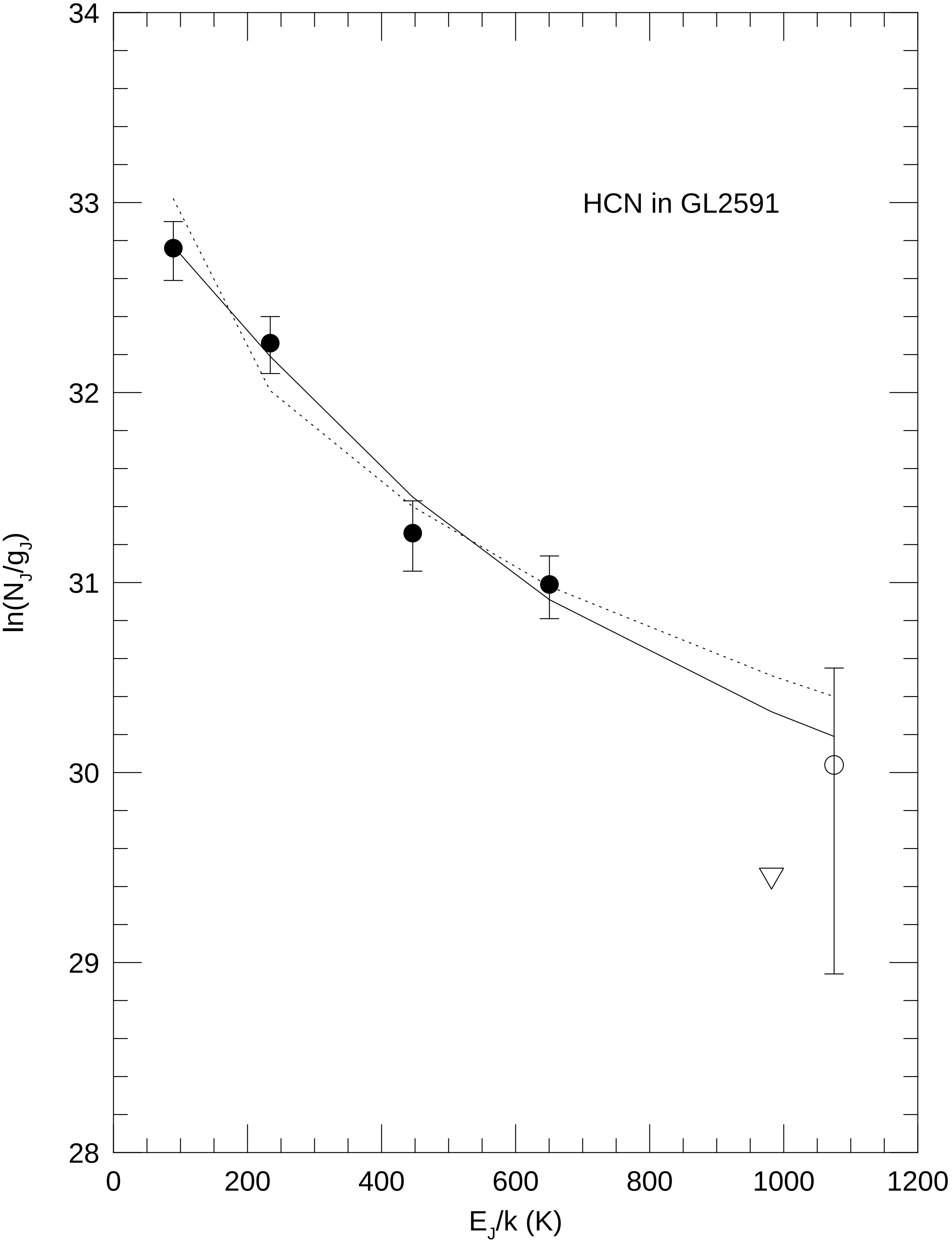

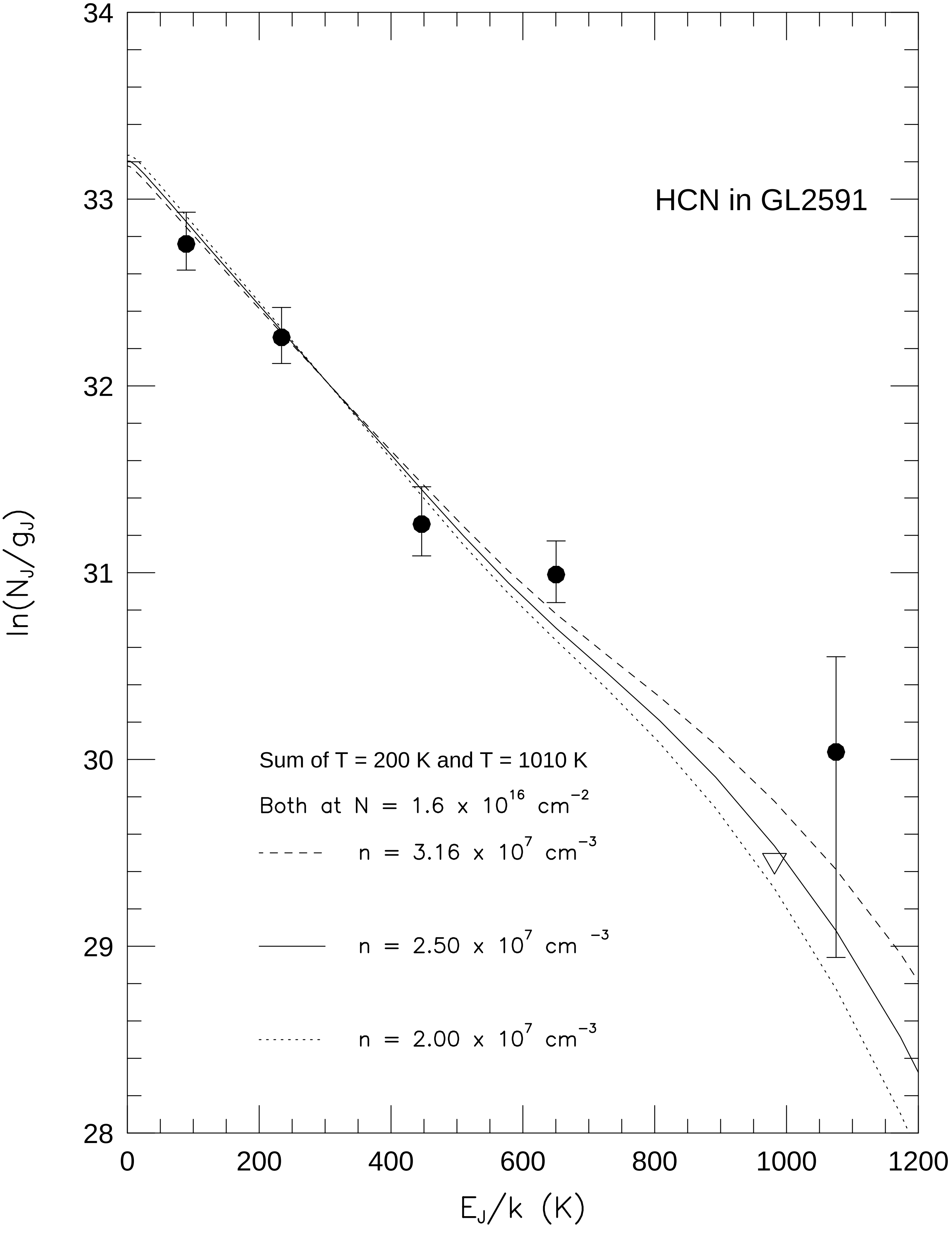

HCN in GL2591
Sum of T = 200 K and T = 1010 K
Both at N = 1.6 x 10^16 cm^-2
n = 3.16 x 10^7 cm^-3
n = 2.50 x 10^7 cm^-3
n = 2.00 x 10^7 cm^-3
ln(N_J/g_J)
E_J/k (K)
34
33
32
31
30
29
28
0
200
400
600
800
1000
1200

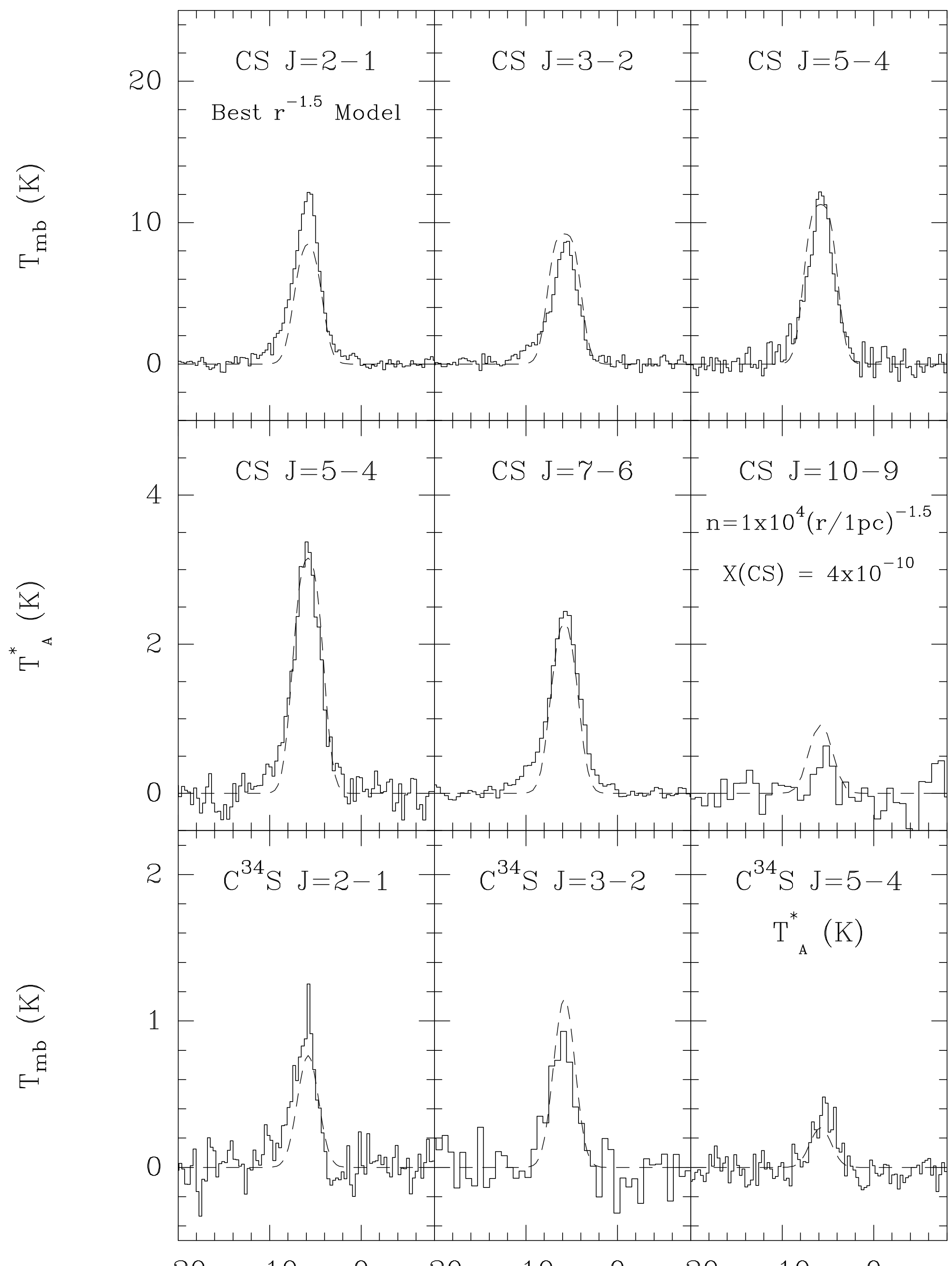

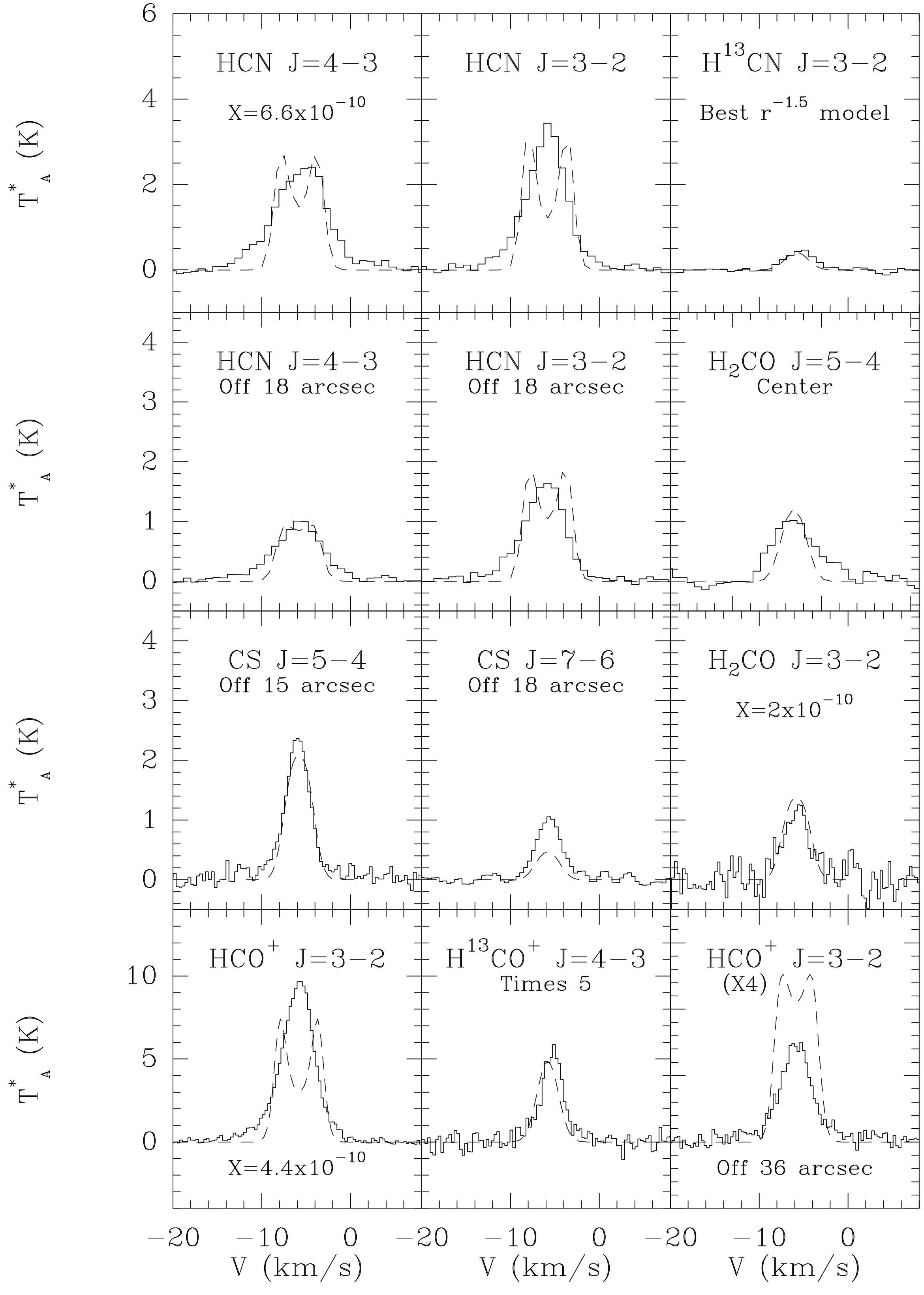

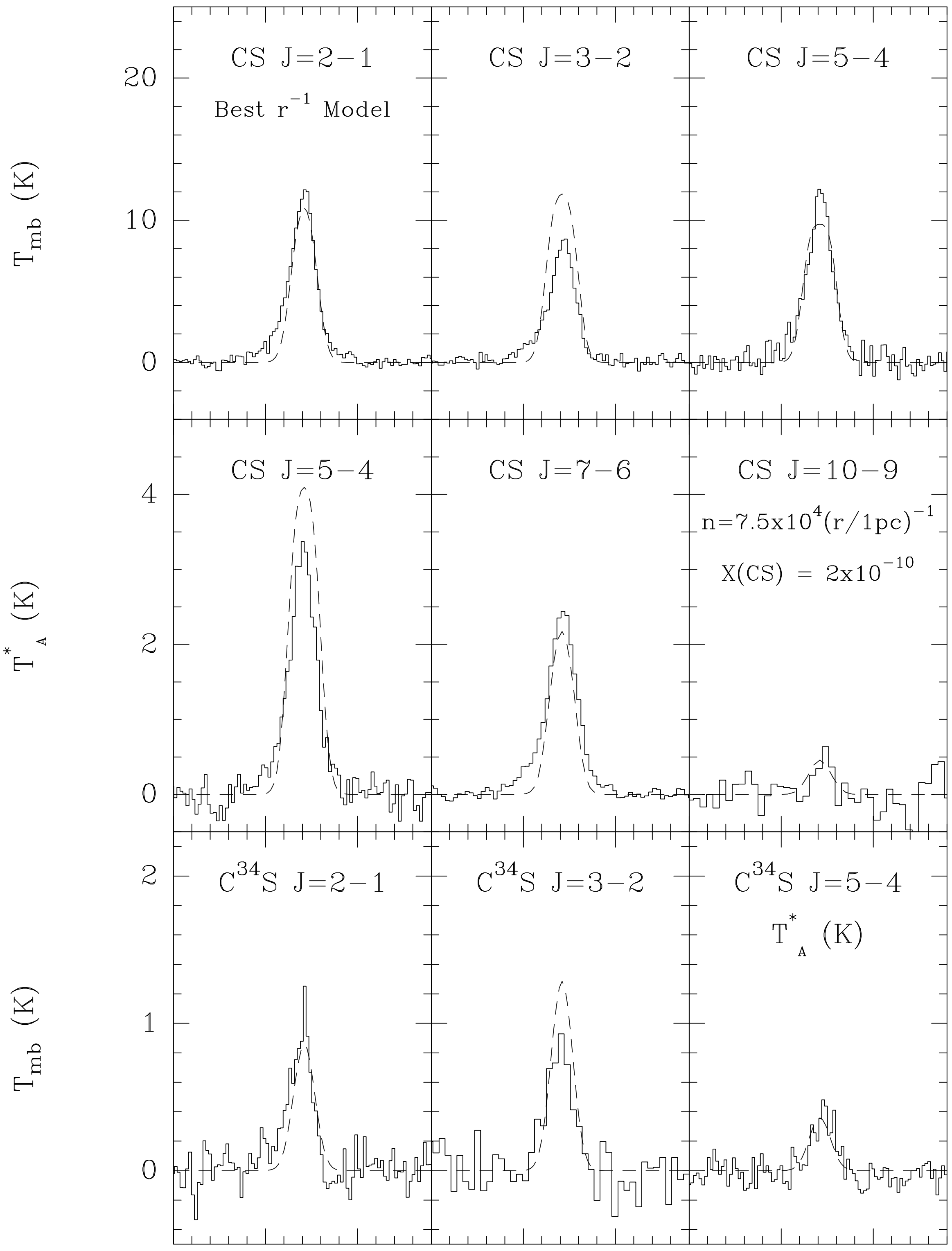

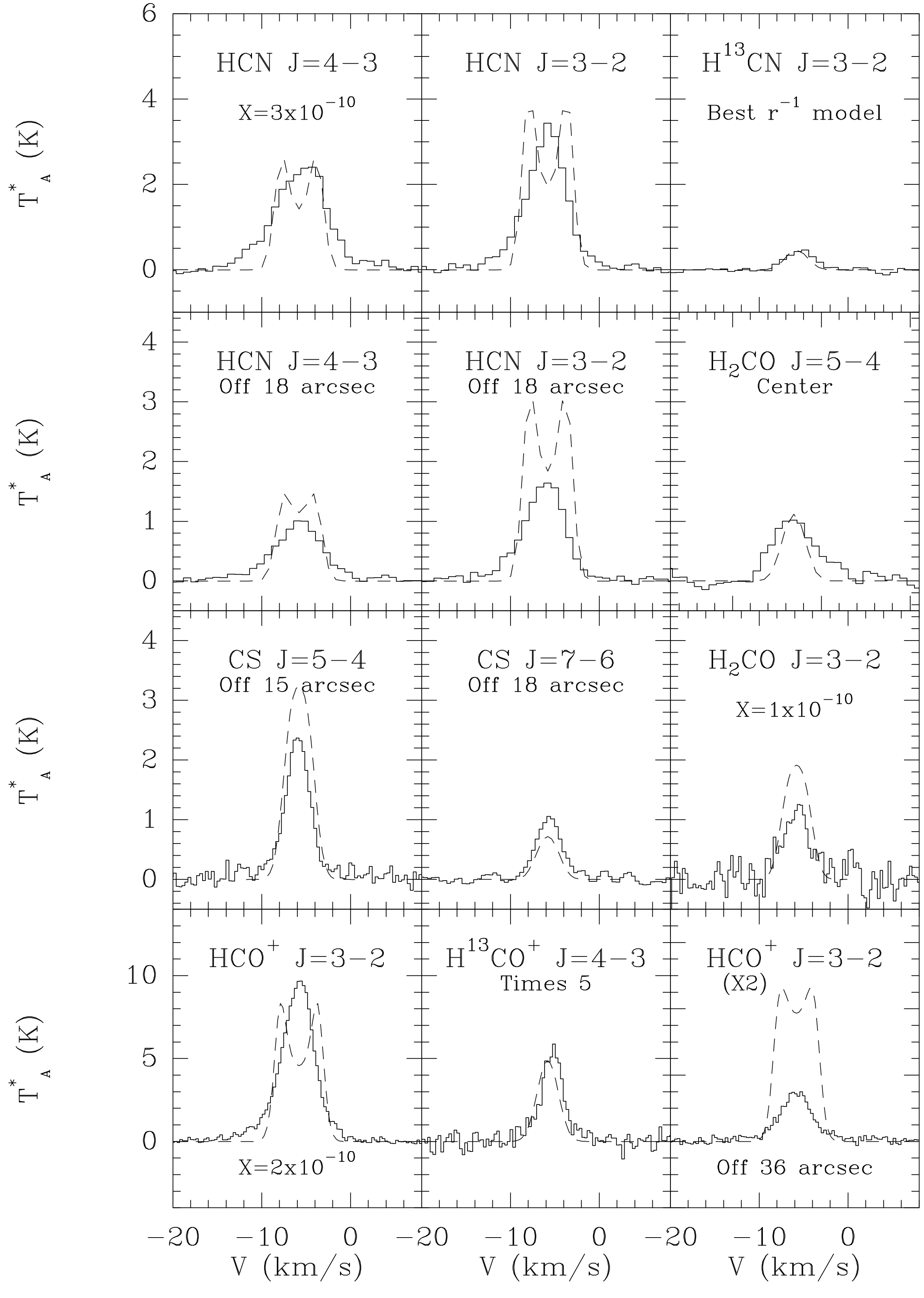

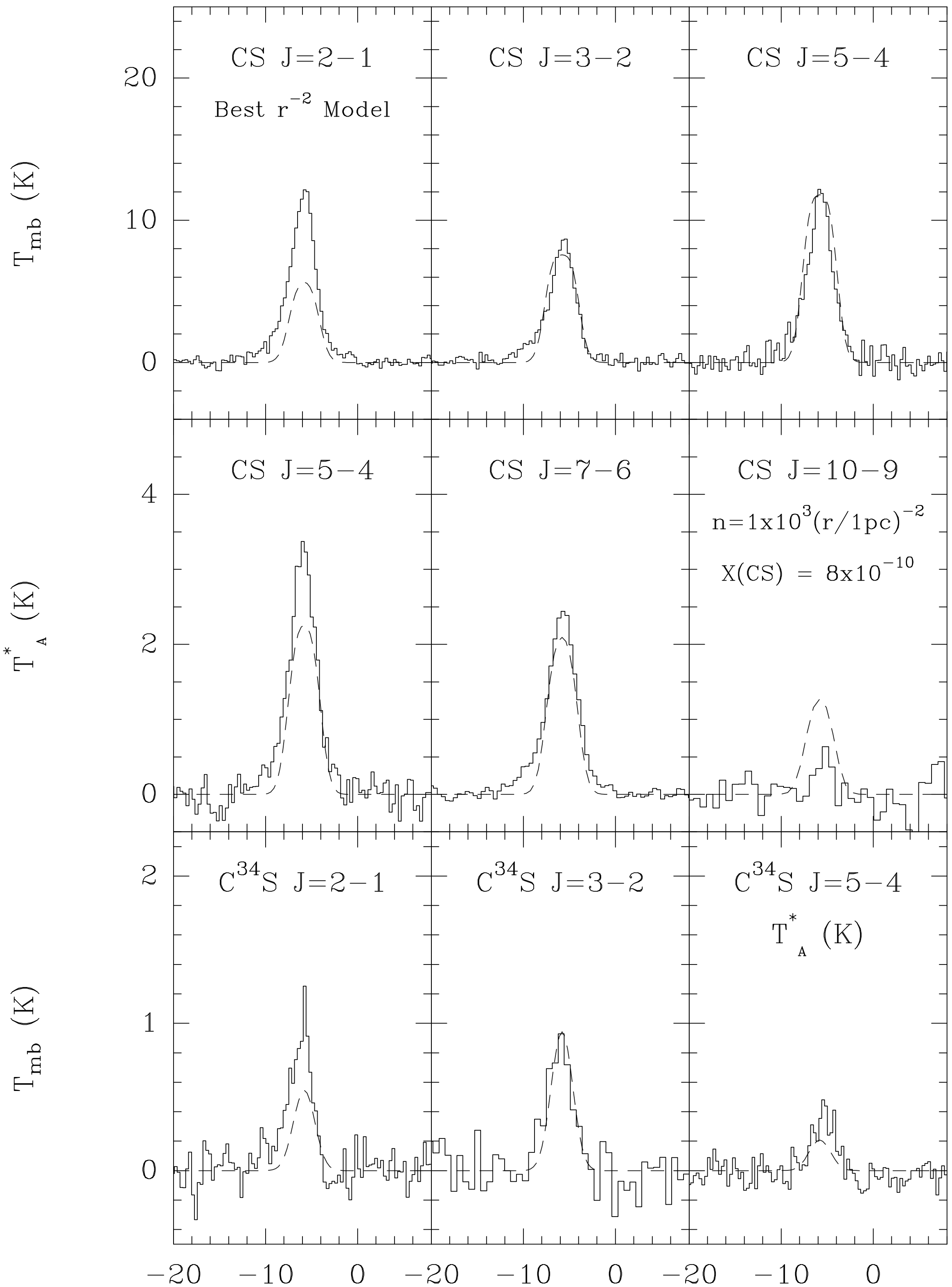

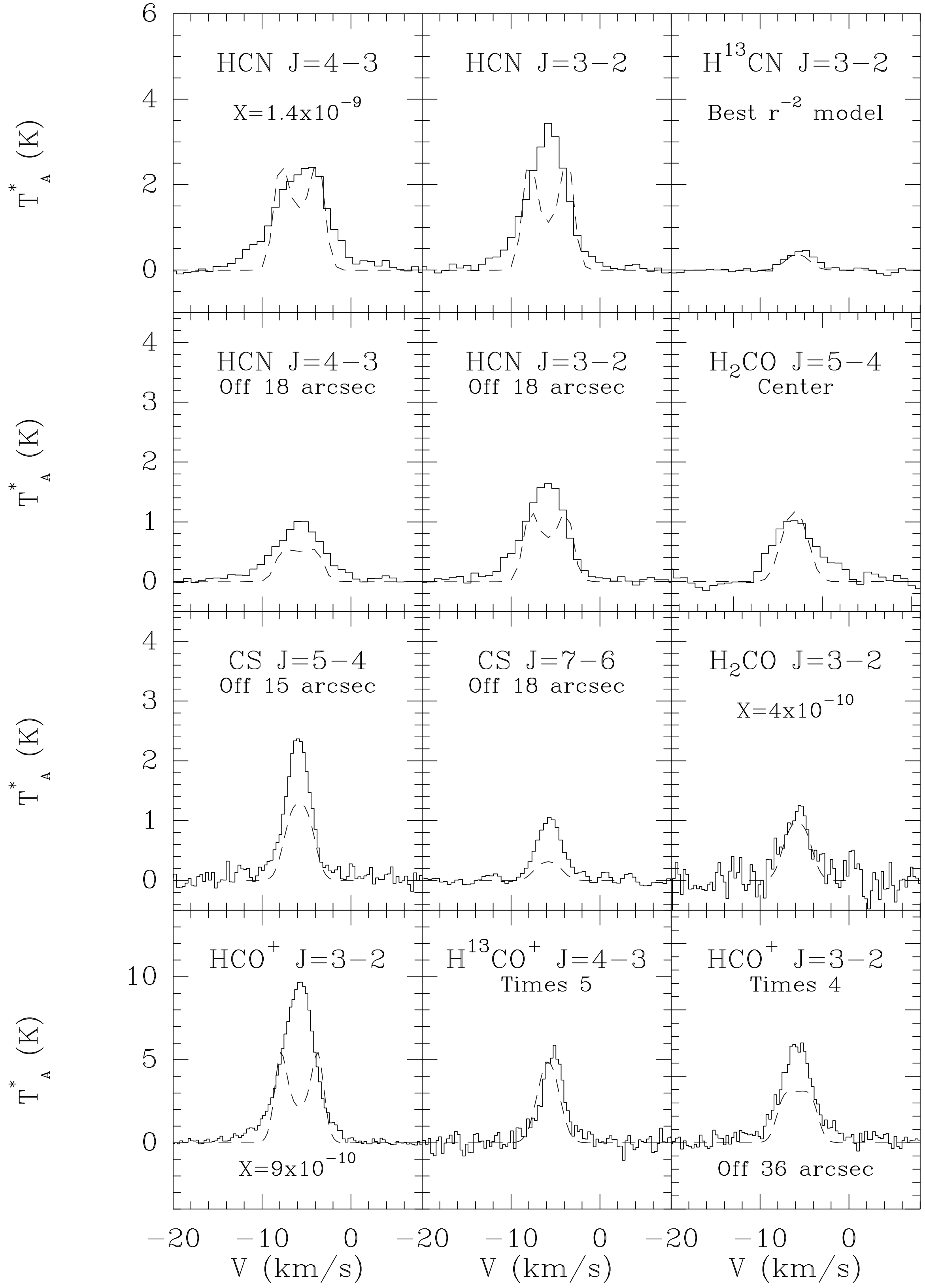